\documentclass[a4paper,11pt]{article}
\pdfoutput=1
\usepackage{jcappub,xcolor}
\usepackage[T1]{fontenc}
\usepackage[shortlabels]{enumitem}

\title{The state of the dark energy equation of state circa 2023}

\author[a,b]{Luis A. Escamilla}
\author[b,1]{William Giar\`e, \note{Corresponding author}}
\author[b]{Eleonora Di Valentino,}
\author[c,d]{Rafael C. Nunes}
\author[e,f]{Sunny Vagnozzi}

\affiliation[a]{Instituto de Ciencias F\'{i}sicas, Universidad Nacional Aut\'{o}noma de M\'{e}xico, Cuernavaca, Morelos, 62210, Mexico}
\affiliation[b]{School of Mathematics and Statistics, University of Sheffield, Hounsfield Road, Sheffield S3 7RH, United Kingdom}
\affiliation[c]{Instituto de F\'{i}sica, Universidade Federal do Rio Grande do Sul, 91501-970 Porto Alegre RS, Brazil}
\affiliation[d]{Divis\~ao de Astrof\'isica, Instituto Nacional de Pesquisas Espaciais (INPE), 12227-010 S\~ao Jos\'e dos Campos SP, Brazil \looseness=-3}
\affiliation[e]{Department of Physics, University of Trento, Via Sommarive 14, 38123 Povo (TN), Italy}
\affiliation[f]{Trento Institute for Fundamental Physics and Applications (TIFPA)-INFN, Via Sommarive 14, 38123 Povo (TN), Italy}

\emailAdd{luis.escamilla@icf.unam.mx}
\emailAdd{w.giare@sheffield.ac.uk}
\emailAdd{e.divalentino@sheffield.ac.uk}
\emailAdd{rafadcnunes@gmail.com}
\emailAdd{sunny.vagnozzi@unitn.it}

\abstract{We critically examine the state of current constraints on the dark energy (DE) equation of state (EoS) $w$. Our study is
motivated by the observation that, while broadly consistent with the cosmological constant value $w=-1$, several independent probes appear to point towards a slightly phantom EoS ($w \sim -1.03$) which, if confirmed, could have important implications for the Hubble tension. We pay attention to the apparent preference for phantom DE from \textit{Planck} Cosmic Microwave Background (CMB) data alone, whose origin we study in detail and attribute to a wide range of (physical and geometrical) effects. We deem the combination of \textit{Planck} CMB, Baryon Acoustic Oscillations, Type Ia Supernovae, and Cosmic Chronometers data to be particularly trustworthy, inferring from this final consensus dataset $w=-1.013^{+0.038}_{-0.043}$, in excellent agreement with the cosmological constant value. Overall, despite a few scattered hints, we find no compelling evidence forcing us away from the cosmological constant (yet).}

\begin{document}
\maketitle
\flushbottom

\section{Introduction}
\label{sec:introduction}

One of the most remarkable discoveries of the past decades, originally determined via observations of distant Type Ia Supernovae (SNeIa)~\cite{SupernovaSearchTeam:1998fmf,SupernovaCosmologyProject:1998vns} and now (indirectly or directly) corroborated by a wide variety of probes~\cite{Sherwin:2011gv,Moresco:2016mzx,Haridasu:2017lma,Rubin:2016iqe,Planck:2018vyg,Yang:2019fjt,Nadathur:2020kvq,Rose:2020shp,DiValentino:2020evt,eBOSS:2020yzd}, is the fact that the current expansion of the Universe is accelerating.\footnote{For important caveats and discussions surrounding this conclusion, see e.g.\ Refs.~\cite{Buchert:1999er,Buchert:2001sa,Buchert:2007ik,Hunt:2008wp,Nielsen:2015pga,Tutusaus:2017ibk,Dam:2017xqs,Colin:2019opb,Desgrange:2019npu,Koksbang:2019cen,Koksbang:2019glb,Heinesen:2022lqs}.} This effect is usually ascribed to a dark energy (DE) component, which adds up to about $70\%$ of the energy budget of the Universe~\cite{Huterer:2017buf}. Within the $\Lambda$CDM cosmological model, the role of DE is played by a cosmological constant $\Lambda$ associated to the zero-point vacuum energy density of quantum fields~\cite{Carroll:2000fy}, whose value expected from theoretical considerations is however in enormous disagreement with observational inferences. Possibly one of the worst fine-tuning problems in theoretical physics, this puzzle goes under the name of ``cosmological constant problem''~\cite{Weinberg:1988cp,Martin:2012bt,Bernardo:2022cck}. This and other considerations have led to the development of a diverse zoo of models for cosmic acceleration which go beyond $\Lambda$, including new (ultra)light fields and modifications to gravity, just to name a few (with no claims as to completeness, see e.g.\ Refs.~\cite{Amendola:1999er,Kamenshchik:2001cp,Capozziello:2002rd,Bento:2002ps,Mangano:2002gg,Farrar:2003uw,Khoury:2003aq,Li:2004rb,Amendola:2006we,Hu:2007nk,Cognola:2007zu,Nojiri:2010pw,Zhang:2011uv,Rinaldi:2014yta,Luongo:2014nld,Rinaldi:2015iza,DeFelice:2016yws,Wang:2016lxa,Josset:2016vrq,Burrage:2016bwy,Sebastiani:2016ras,Nojiri:2017ncd,Burrage:2017qrf,Capozziello:2017buj,Benisty:2018qed,Casalino:2018tcd,Yang:2018euj,Saridakis:2018unr,Visinelli:2018utg,Langlois:2018dxi,Benisty:2018oyy,Boshkayev:2019qcx,Heckman:2019dsj,DAgostino:2019wko,Mukhopadhyay:2019wrw,Mukhopadhyay:2019cai,Mukhopadhyay:2019jla,Vagnozzi:2019kvw,Akarsu:2019hmw,Saridakis:2020zol,Ruchika:2020avj,Odintsov:2020zct,Odintsov:2020vjb,Oikonomou:2020qah,Oikonomou:2020oex,Vagnozzi:2021quy,Solanki:2021qni,Saridakis:2021qxb,Arora:2021tuh,Capozziello:2022jbw,Narawade:2022jeg,DAgostino:2022fcx,Oikonomou:2022wuk,Belfiglio:2022egm,Luciano:2022hhy,Kadam:2022yrj,Ong:2022wrs,Bernui:2023byc,Luciano:2023wtx,Giani:2023tai,Belfiglio:2023rxb,Frion:2023xwq,Adil:2023ara}).

The final word on which DE model(s) best describe our Universe ultimately rests with observational (cosmological) data. Besides the intrinsic value in understanding what 70\% of the Universe is made of, the importance of this quest is reinforced by the growing ``Hubble tension''~\cite{Verde:2019ivm,DiValentino:2021izs,Perivolaropoulos:2021jda,Abdalla:2022yfr,Hu:2023jqc}, i.e.\ the disagreement between a number of independent probes of the Hubble rate $H_0$, which could ultimately have important implications for the nature of (late) DE (see e.g.\ Refs.~\cite{DiValentino:2016hlg,Mortsell:2018mfj,Dutta:2018vmq,Li:2019yem,Vagnozzi:2019ezj,Visinelli:2019qqu,DiValentino:2019ffd,Dutta:2019pio,DiValentino:2019jae,Zumalacarregui:2020cjh,Alestas:2020mvb,DiValentino:2020naf,Alestas:2020zol,Kumar:2021eev,Vagnozzi:2021tjv,Bag:2021cqm,Theodoropoulos:2021hkk,Alestas:2021luu,Sen:2021wld,Roy:2022fif,Heisenberg:2022lob,Chudaykin:2022rnl,Akarsu:2022typ,Santos:2022atq,Schiavone:2022wvq,Ben-Dayan:2023rgt,Ballardini:2023mzm,deCruzPerez:2023wzd,Zhai:2023yny,Adil:2023exv,Montani:2023xpd,Akarsu:2023mfb,Vagnozzi:2023nrq,Avsajanishvili:2023jcl,Giani:2023aor,Lazkoz:2023oqc,Akarsu:2024qiq,Benisty:2024lmj,Moshafi:2024guo}).\footnote{The qualifier ``late'' is helpful to distinguish the DE component we will be interested in throughout the rest of this paper, from a more speculative ``early'' DE component, operative mostly around recombination and invoked to solve the Hubble tension~\cite{Poulin:2018cxd,Niedermann:2019olb,Sakstein:2019fmf,Kamionkowski:2022pkx,Poulin:2023lkg}.} One of the key properties of DE is its equation of state (EoS) $w \equiv P_x/\rho_x$, i.e.\ the ratio between the DE pressure $P_x$ and energy density $\rho_x$, which takes the value $w=-1$ for $\Lambda$. Unsurprisingly, a sub-percent level determination of $w$ is among the key scientific goals of a number of upcoming and recently ongoing cosmological surveys~\cite{LSST:2008ijt,J-PAS:2014hgg,Spergel:2015sza,Amendola:2016saw,CMB-S4:2016ple,DESI:2016fyo,SimonsObservatory:2018koc,Slosar:2019flp,SimonsObservatory:2019qwx,Bonoli:2020ciz}, whereas the question of what cosmological data has to say about $w$ is one which has been tackled in a several earlier works, albeit not always reaching definitive conclusions~\cite{Bean:2001xy,Hannestad:2002ur,Said:2013jxa,Shafer:2013pxa,Zhang:2015uhk,Xu:2016grp,Wang:2016tsz,Vagnozzi:2017ovm,Zhang:2017rbg,Feng:2017mfs,Wang:2018ahw,Sprenger:2018tdb,Poulin:2018zxs,RoyChoudhury:2018vnm,Wang:2019acf,RoyChoudhury:2019hls,Zhang:2021yof,Colgain:2021pmf,Teng:2021cvy,Krishnan:2021dyb,Nunes:2021ipq,Bernardo:2021cxi,Wang:2022xdw,Koussour:2022jyk,Bernardo:2022pyz,Narawade:2022cgb,Hou:2022rvk,Kumar:2023bqj,Bhagat:2023ych,Mussatayeva:2023aoa} -- in passing, we note that our title is inspired by the earlier seminal Ref.~\cite{Melchiorri:2002ux}, written over two decades ago.

Aside from being the EoS of $\Lambda$, $w=-1$ plays a very important demarcation role, as it separates two widely different physical regimes. In fact, a component with $w<-1$ and positive energy density $\rho_x>0$ would violate the null energy condition (and therefore all other energy conditions), which stipulates that $T_{\mu\nu}k^{\mu}k^{\nu} \geq 0$, where $T_{\mu\nu}$ is the stress-energy tensor of the component and $k^{\mu}$ is an arbitrary future-pointing null vector. A component with $w<-1$ is usually referred to as ``phantom'', and from the point of view of the cosmological evolution it is well-known that an Universe dominated by phantom DE would terminate its existence in a so-called Big Rip, i.e.\ the dissociation of bound systems due to the phantom energy density becoming infinite in a finite amount of time~\cite{Caldwell:2003vq} (see also Refs.~\cite{Nojiri:2005sx,Frampton:2011sp,Astashenok:2012tv,Odintsov:2015zza,Odintsov:2018zai}). On the other hand, a component with $w>-1$ (and $w<-1/3$ to ensure cosmic acceleration) is sometimes referred to as lying in the ``quintessence-like'' regime, a reference to the simplest quintessence models of DE~\cite{Wetterich:1987fm,Ratra:1987rm,Wetterich:1994bg,Caldwell:1997ii,Sahni:2002kh}.\footnote{Note, however, that the simplest quintessence models featuring a single scalar field with a canonical kinetic term, minimally coupled to gravity, and without higher derivative operators, are at face value disfavored observationally, as they worsen the Hubble tension~\cite{Vagnozzi:2018jhn,OColgain:2018czj,Colgain:2019joh,Banerjee:2020xcn,Heisenberg:2022gqk}.} Clearly, a high-precision determination of the value of $w$ would have tremendous implications in terms of understanding the ultimate fate of the Universe, but the ramifications for fundamental physics and model-building would also be momentous -- it is no exaggeration to say that a compelling confirmation of DE's nature being phantom would force a large fraction of the particle physics and cosmology communities back to the blackboard~\cite{Carroll:2003st,Vikman:2004dc,Carroll:2004hc,Hu:2004kh,Saridakis:2008fy,Leon:2009dt,Deffayet:2010qz,Sawicki:2012pz,Nojiri:2013ru,Maggiore:2013mea,Oikonomou:2014gsa,Nunes:2015rea,Nojiri:2015fia,Ludwick:2017tox,Odintsov:2018obx}.

The recent launch of a number of so-called ``Stage IV'' surveys, and the planned (imminent) launch of a number of others, all of whom share among their goals a precise determination of $w$ and thereby of the nature of DE, motivates us to provide a re-appraisal of the status of current (at the time of writing, mid-2023) cosmological inferences of $w$. It is often stated that current data is remarkably consistent with the cosmological constant picture, where $w=-1$. This is true in a broad sense, yet it is undeniable that there are a number of indications that things might not be so simple. Digging a bit deeper, one can indeed find hints for deviations from $w=-1$ inferred from current data (particularly in the phantom direction, pointing towards $w \sim -1.03$), which we feel may have somewhat been brushed under the carpet or not given sufficient attention. It is our goal in this paper to look in more detail into the latest constraints on $w$, understand whether the hints mentioned earlier are indeed present and what is driving them, and provide a complete picture of the current state of the DE EoS.

The rest of this paper is then organized as follows. In Sec.~\ref{sec:deeos} we briefly review how the DE EoS enters into the main equations used to describe the evolution of the background expansion and cosmological perturbations, and even more briefly present the aforementioned hints for potential deviations from $w=-1$ in current data. In Sec.~\ref{sec:data} we discuss the datasets and analysis methodology which will be used in the remainder of our paper. The results of our analysis and a thorough physical interpretation thereof are presented in Sec.~\ref{sec:results}. Finally, in Sec.~\ref{sec:conclusions} we draw concluding remarks. Note that throughout our paper, we consider a barotropic DE component with EoS constant in time, i.e.\ $w(z)=w$.

\section{The dark energy equation of state}
\label{sec:deeos}

We begin by briefly reviewing how the DE EoS enters the basic equations relevant for computing the cosmological observables considered here. At the background level, the DE energy density $\rho_x(z)$ evolves as a function of redshift as:
\begin{eqnarray}
\rho_x(z)=\rho_{x,0}(1+z)^{3(1+w)}\,,
\label{eq:rhox}
\end{eqnarray}
where $\rho_{x,0}$ indicates the present-time DE energy density. Clearly, if $w=-1$ as in the case of a cosmological constant, $\rho_x(z)$ is indeed a constant. This in turn affects the expansion rate of the Universe via the Friedmann equation:
\begin{eqnarray}
H^2(a)=\frac{8\pi G}{3}\rho - \frac{kc^2}{a^2},
\label{eq:friedmann1}
\end{eqnarray}
where $H(a)$ is the Hubble Parameter as a function of scale factor, $G$ is Newton's constant, $\rho$ is the energy-density (which includes the DE energy density $\rho_x$), and $k$ is the curvature. It is straightforward to see that by modifying $w$, one modifies $\rho_x$ and consequently $H(a)$.

Besides altering the evolution of the cosmological background, a DE EoS $w \neq -1$ also affects the evolution of perturbations. To study these, we work in synchronous gauge~\cite{Lifshitz:1945du,Ma:1995ey}, where the perturbed Friedmann-Lema\^{i}tre-Robertson-Walker line element is given by:
\begin{eqnarray}
ds^2 = a^2(\eta) \left [ d\eta^2+(\delta_{ij}+h_{ij}) \right ] \,,
\label{eq:ds2}
\end{eqnarray}
with $a$ and $\eta$ denoting the scale factor and conformal time respectively, whereas we denote by $h \equiv h_{ii}$ the trace of the synchronous gauge metric perturbation $h_{ij}$. Moving to Fourier space, we further denote by $\delta_x$ and $\theta_x$ the DE density contrast and velocity perturbation respectively. At linear order in perturbations, and further assuming that the DE sound speed squared $c_{s,x}^2$ (given by the ratio between the DE pressure and density perturbations in the DE rest frame) is $c_{s,x}^2=1$, as expected for the simplest DE models based on a single light, minimally coupled scalar
field, with a canonical kinetic term, and in the absence of higher order operators, the evolution of $\delta_x$ and $\theta_x$ is governed by the following equations:
\begin{align}
\label{eq:deltax}
&\dot{\delta}_x = -(1+w) \left ( \theta_x+\frac{\dot{h}}{2} \right ) -3{\cal H}(1-w) \left [ \delta_x+\frac{3{\cal H}(1+w)\theta_x}{k^2} \right ]\,, \\
\label{eq:thetax}
&\dot{\theta}_x = 2{\cal H}\theta_x+\frac{k^2}{1+w}\delta_x\,,
\end{align}
where we have also set the DE adiabatic sound speed squared $c_{a,x}^2=w$, as expected for a non-interacting DE component whose EoS is constant in time.\footnote{See e.g.\ Sec.~2.1 of Ref.~\cite{Ferlito:2022mok} for a recent more detailed discussion on the range of validity of these assumptions.}

Note that, although at first glance Eqs.~(\ref{eq:deltax},\ref{eq:thetax}) diverge in the cosmological constant limit $w \to -1$, what actually happens is that the perturbation equations become irrelevant, as $\Lambda$ is perfectly smooth and cannot support perturbations. Moreover, as we are considering a DE EoS $w$ which is constant in time, always either in the quintessence-like regime ($w>-1$) or in the phantom regime ($w<-1$), but which cannot cross from one to another, we do not need to make use of the parameterized post-Friedmann approach~\cite{Hu:2007pj,Fang:2008sn,Li:2014eha} when treating DE perturbations.

The value of $w$ can be inferred through a variety of cosmological probes, including those which are (directly or indirectly) sensitive to its effect on the background expansion and thereby on distances, as well as others more sensitive to its effect on the growth of structure. Intriguingly, a diverse range of recent, independent probes, taken at face value, all seem to point towards a slightly phantom DE component, with $w \sim -1.03$. One such probe are Cosmic Microwave Background (CMB) measurements from the \textit{Planck} satellite. However, due to the so-called geometrical degeneracy, various combinations of these parameters can be arranged so as to keep the acoustic angular scale $\theta_s$ fixed. This is given by the ratio of the comoving sound horizon at recombination to the comoving distance to last-scattering -- if both change by the same amount, the acoustic angular scale remains unchanged. As a result, these measurements alone cannot, by construction, provide a strong constraint on $w$, unless perturbation-level effects are included. Nevertheless, they appear to point towards a $\simeq 2\sigma$ preference for phantom DE~\cite{Planck:2018vyg}, whose origin we shall attempt to investigate (among other things) in this paper.\footnote{It is important to mention that the preference for phantom DE found in Ref.~\cite{Planck:2018vyg} is slightly below $2\sigma$ for certain dataset combinations, e.g.\ for TT+TE+EE+lowE, TT+TE+EE+lowE+lensing, and when BAO data are included. Nevertheless, the preference for $w<-1$ persists even in these cases.}

The earlier weak indication for phantom DE persists when \textit{Planck} data are combined with other external datasets which are able to break the geometrical degeneracy, such as Baryon Acoustic Oscillation (BAO) measurements, which once combined with \textit{Planck} indicate a central value of $w \sim -1.04$ (see Tab.~4 of Ref.~\cite{eBOSS:2020yzd}).\footnote{An interesting point, which yet has not been discussed in depth in the literature, can be seen in Tab.~4 of Ref.~\cite{eBOSS:2020yzd}: when inferring $w$ from BAO data alone, one finds $w=-0.69 \pm 0.15$, deep in the quintessence-like regime. Upon closer inspection it is apparent that, despite BAO data alone favoring the quintessence-like regime, $H_0$ is completely unconstrained. For this reason, we refrain from further interpreting this result.} Combining \textit{Planck} with full-shape (FS) galaxy power spectrum measurements (which of course also include BAO information) leads to even stronger hints for $w<-1$, with varying strength depending on the underlying analysis assumption and additional external datasets included (see e.g.\ Tabs.~1 and~7 of Ref.~\cite{DAmico:2020kxu};  Tab.~I of Ref.~\cite{Chudaykin:2020ghx}; Tab.~ of Ref.~\cite{Brieden:2022lsd}; Tabs.~3 and~5 of Ref.~\cite{Carrilho:2022mon}; and Tab.~2 of Ref.~\cite{Semenaite:2022unt}, for examples of recent analyses in these directions, all inferring $w<-1$, with central values ranging between $-1.17$ and $-1.03$).

Other slight ($\lesssim 2\sigma$) hints for phantom DE have been observed when combining \textit{Planck} with weak lensing (WL) data. In Ref.~\cite{DES:2018ufa} the $w_0w_a$ parametrization for the DE EoS [$w(a)=w_0+(1-a)w_a$] was used to infer $w_0=-0.96^{+0.10}_{-0.08}$ and $w_a=-0.31^{+0.38}_{-0.52}$ (therefore a component evolving towards the phantom regime in the past) using DES data, as shown in Tab.~III of Ref.~\cite{DES:2018ufa}; in Ref.~\cite{KiDS:2020ghu} using the $w$CDM parametrization, a value of $w=-1.05^{+0.21}_{-0.26}$ was obtained from KiDS data, although due to the large uncertainties this is largely in agreement with $\Lambda$CDM. Finally, using the $w_0w_a$ parametrization, Ref.~\cite{DES:2022ccp} finds best-fit values $w_0=-0.94^{+0.08}_{-0.08}$ and $w_a=-0.45^{+0.36}_{-0.28}$ (therefore a component evolving towards the phantom regime in the past), as reported in Tab.~IV of Ref.~\cite{DES:2022ccp}. However, it is important to exercise caution as some of these results are based on so-called $3 \times 2$pt analyses and therefore also use galaxy clustering data, so the results are not completely independent from those involving BAO data reported earlier.

Similar hints were also present when combining \textit{Planck} data with the \textit{Pantheon} SNeIa sample (Tabs.~12,~13 and~14 of Ref.~\cite{Pan-STARRS1:2017jku}), although with the \textit{PantheonPlus} sample such a hint has disappeared, with $w>-1$ preferred (Tab.~3 of Ref.~\cite{Brout:2022vxf}). While the error bars are larger than all the previously mentioned cases, combinations of \textit{Planck} and cosmic chronometers data, which we will discuss in more detail later, also prefer phantom DE (Tabs.~1 and~3 of Ref.~\cite{Moresco:2016nqq} and Tab.~4 of Ref.~\cite{Vagnozzi:2020dfn}). Unsurprisingly, such a preference is obviously obtained when combining \textit{Planck} data with a local prior on the Hubble constant $H_0$, given the ongoing Hubble tension and the direction of the $H_0$-$w$ degeneracy at the level of the Friedmann equation, which tends to push $w$ to lower values when $H_0$ is increased: see Tab.~5--8 of Ref.~\cite{DiValentino:2020vnx} and Tabs.~II--VI in Ref.~\cite{Yang:2021flj} for concrete examples in the context of analyses of varying degree of conservativeness.

Finally, hints for phantom DE with varying degree of robustness have been obtained when considering a number of late-time so-called ``emerging probes'' beyond the standard BAO, SNeIa, and WL ones, many of which have been recently presented in detail in the review of Ref.~\cite{Moresco:2022phi}. For instance, preferences for phantom DE have been reported using quasar data~\cite{Bargiacchi:2021hdp} (Tab. 2), both alone and in combination with BAO and SNeIa, gamma-ray bursts (see Tab.~5 of Ref.~\cite{Moresco:2022phi}), cluster strong lensing cosmography (Fig. 7 of Ref.~\cite{Grillo:2020yvj}), HII starburst galaxies (Tabs.~1 and~2 of Ref.~\cite{Cao:2021cix}), and so on. Although all these probes come with varying degree of credibility and associated uncertainties, when looking at the big picture there appears to be an overall slight preference for phantom DE. When combining \textit{Planck} measurements with the most credible low-redshift measurements, one appears to be drawn to the conclusion that DE is slightly phantom, with $w$ just slightly below $-1$ (indicatively, between $-1.05$ and $-1.03$). While such a component is only very weakly in the phantom regime, if correct and confirmed, such a result would nonetheless have very important repercussions for fundamental physics: for this reason, it is our goal in this paper to look in more detail into the latest constraints on $w$, with an eye to these hints for phantom DE.

\section{Datasets and methodology}
\label{sec:data}

\begin{table}[!ht]
	\begin{center}
		\footnotesize
		\renewcommand{\arraystretch}{1.5}
		\begin{tabular}{l@{\hspace{0. cm}}@{\hspace{2 cm}} c}
			\hline\hline
			\textbf{Parameter} & \textbf{Prior} \\
			\hline\hline
			$\Omega_{\rm b} h^2$ & $[0.005\,,\,0.1]$ \\
			$\Omega_{\rm c} h^2$ & $[0.005\,,\,0.99]$ \\
			$\tau$ & $[0.01, 0.8]$ \\
			$100\,\theta_s$ & $[0.5\,,\,10]$ \\
			$\log(10^{10}A_{\rm S})$ & $[1.61\,,\,3.91]$ \\
			$n_{\rm s}$ & $[0.8\,,\, 1.2]$ \\
			$w$ & $[-3\,,\,1]$ \\
			\hline\hline
		\end{tabular}
		\caption{Ranges for the flat prior distributions imposed on cosmological parameters in our analyses.}
		\label{table:priors}
	\end{center}
\end{table}

In order to determine constraints on the dark energy equation of state $w$ we will perform, from scratch, a parameter inference analysis within the $w$CDM model using a number of different datasets, both alone and in various combinations. Specifically, we consider:
\begin{itemize}
	\item Measurements of CMB temperature anisotropy and polarization power spectra, their cross-spectra, and the lensing power spectrum reconstructed from the temperature 4-point correlation function, all from the \textit{Planck} 2018 legacy data release, using the official \textit{Planck} likelihoods (to be more specific the \textit{plik} TTTEEE+\texttt{lowE}+lensing) ~\cite{Planck:2018nkj}. 
    We refer to this dataset combination as \textbf{\textit{CMB}}, and the official likelihood and data products can be obtained in the public repository~\cite{Planck_archive}, with the particular file being~\cite{Planck_file}. 
    One of our aims in this work will be to identify the physical origin of constraints from CMB data, and more precisely which scales/multipole ranges are driving our constraints and preference for phantom DE. In order to address this issue, we will also perform multipole splits, i.e.\ only looking at data within a certain $\ell$ range (an approach followed earlier, among others, in Ref.~\cite{Planck:2016tof}). These choices of range may or may not entail the use of the low-$\ell$ EE-only ($2 \leq \ell \leq 29$) \texttt{SimAll} likelihood, which we refer to as \texttt{lowE}. When performing such multipole splits, we will be explicit about the $\ell$ range which is being used, and whether or not \texttt{lowE} data is included.
	\item Baryon Acoustic Oscillation (BAO) and Redshift-Space Distortions (RSD) measurements from the completed SDSS-IV eBOSS survey. These include isotropic and anisotropic distance and expansion rate measurements, and measurements of $f\sigma_8$, and are summarized in Tab.~3 of Ref.~\cite{eBOSS:2020yzd}, whereas the datasets can be found in the public repository~\cite{BAO_archive} (for the \texttt{CosmoMC} implementation of the likelihoods refer to the public repository~\cite{BAO_github}). We collectively denote this set of measurements as \textbf{\textit{BAO}}.
	\item The \textit{PantheonPlus} SNeIa sample~\cite{Scolnic:2021amr}, consisting of 1701 light curves for 1550 distinct SNeIa in the redshift range $0.001<z<2.26$. In all but one case we will consider the uncalibrated \textit{PantheonPlus} SNeIa sample, whereas in the remaining case we will also consider SH0ES Cepheid host distances, which can be used to calibrate the SNeIa sample\cite{Riess:2021jrx}. We refer to the uncalibrated sample as \textbf{\textit{SN}}, whereas the SH0ES calibration (when included) is referred to as \textbf{\textit{SH0ES}}. The corresponding likelihood can be found in the public repository~\cite{Pantheon_github}.
	\item Measurements of the expansion rate $H(z)$ from the relative ages of massive, early-time, passively-evolving galaxies, known as cosmic chronometers (CC)~\cite{Jimenez:2001gg}. We use 15 CC measurements in the range $0.179<z<1.965$, compiled in Refs.~\cite{Moresco:2012by,Moresco:2015cya,Moresco:2016mzx}. Although in principle more than 30 CC measurements are available, we choose to restrict our analysis to this subset, for which a full estimate of non-diagonal terms in the covariance matrix, and systematics contributions to the latter, is available, and obtained following Refs.~\cite{Moresco:2018xdr,Moresco:2020fbm}. We also choose to exclude some of the earlier measurements due to the concerns expressed in Ref.~\cite{Kjerrgren:2021zuo}, against which our sample is instead safe. Finally, we note that including the additional available CC measurements is not expected to drastically alter our conclusions, as our sample already contains some of the measurements with the smallest (and at the same time most reliable) uncertainties. We refer to the set of 15 measurements we adopted as \textbf{\textit{CC}}, and the corresponding data is available in the public repository~\cite{CC_github}.
	\item Full-shape galaxy power spectrum measurements from the BOSS DR12 galaxy sample. In particular, the dataset we use includes measurements of the lowest-order galaxy power spectrum multipoles (P0,P2,P4)~\cite{Ivanov:2019pdj,Philcox:2020vvt}, the real-space power spectrum proxy statistic $Q_0$~\cite{Ivanov:2021fbu} and the bispectrum monopole~\cite{Philcox:2021kcw}. The modelling of these observables is based on the so-called Effective Field Theory of Large-Scale Structure (EFTofLSS)~\cite{Baumann:2010tm,Carrasco:2012cv,Cabass:2022avo}, a symmetry-driven model for the mildly non-linear clustering of biased tracers of the large-scale structure such as galaxies, which integrates out the complex and poorly-known details of short-scale (UV) physics. We refer to this dataset as \textbf{\textit{FS}}, and the likelihood is available in the public repository~\cite{FS_github}.
\end{itemize}
Given the large number of datasets used in our analysis, and their being unavailable in a single repository, we employed two public Markov Chain Monte Carlo (MCMC) samplers to perform the parameter inference: \texttt{Cobaya}~\cite{Torrado:2020dgo} in conjunction with the Boltzmann solver \texttt{CAMB}~\cite{Lewis:1999bs} for almost every run, and \texttt{Montepython}~\cite{Brinckmann:2018cvx,Audren:2012wb} in conjunction with the Boltzmann solver \texttt{CLASS-PT}~\cite{Chudaykin:2020aoj}, itself a modified version of \texttt{CLASS}~\cite{Blas:2011rf}, for the runs involving the FS dataset.

We assess the convergence of our chains using the Gelman-Rubin parameter $R-1$~\cite{Gelman:1992zz}, setting the threshold $R-1<0.02$ for our chains to be deemed converged. Model-wise, we consider the 7-parameter $w$CDM model. Specifically, the first six parameter we consider are the standard $\Lambda$CDM ones: the physical baryon density $\Omega_bh^2$, the physical dark matter density parameter $\Omega_ch^2$, the optical depth $\tau$, the angular size of the sound horizon at recombination $\theta_s$, the amplitude of primordial scalar perturbations $\log{(10^{10} A_s)}$, and the scalar spectral index $n_s$. The seventh parameter is the DE EoS $w$. We set a wide flat priors on $w \in [-3;1]$, which allows it to enter both the quintessence-like and phantom regimes, and verify a posteriori that our results are not affected by the choice of lower and upper prior boundaries\footnote{Note that in any case $w<-1/3$ is required in order for the Universe to undergo accelerated expansion.}. Besides the cosmological parameters discussed above, we explicitly vary and marginalize over all the nuisance parameters required to analyze the datasets in question (we refer the reader to the original papers for detailed discussions on these nuisance parameters and their treatment). The prior ranges we adopted for the seven cosmological parameters are summarized in ~\autoref{table:priors}. In this study, rather than conducting an extensive model comparison analysis of $w$CDM versus $\Lambda$CDM, our  goal is to examine the behavior of the $w$CDM model, and more specifically the DE EoS parameter $w$, when confronted against the targeted dataset combinations. Therefore, Bayesian evidences will not be computed.

\begin{table*}[!ht]
	\footnotesize
	\renewcommand{\arraystretch}{1.7}
	\resizebox{\textwidth}{!}{
		\begin{tabular}{ccc|ccc} 
			\centering
			\boldmath{$\ell_{\rm{max}}$} \textbf{(with \texttt{lowE})} & \boldmath{$w$} & \boldmath{$H_0$}~\textbf{[km/s/Mpc]} & \boldmath{$\ell_{\rm{max}}$} \textbf{(without \texttt{lowE})} & \boldmath{$w$} & \boldmath{$H_0$}~\textbf{[km/s/Mpc]}\\
			\hline\hline
			2500 & $-1.57^{+0.16}_{-0.36}$ ($-1.57^{+0.53}_{-0.42}$) & $>82.1 (>69.3)$ & 2500 & $-1.47^{+0.20}_{-0.40}$ ($-1.47^{+0.57}_{-0.47}$) & $>78.7 (>65.9)$ \\
			
			2200 & $-1.58^{+0.16}_{-0.35}$ ($-1.58^{+0.54}_{-0.42}$) & $>82.4 (>69.3)$ & 2200 & $-1.48^{+0.19}_{-0.39}$ ($-1.48^{+0.56}_{-0.47}$) & $>79.1 (>66.4)$ \\
			
			2000 & $-1.57^{+0.16}_{-0.36}$ ($-1.57^{+0.54}_{-0.42}$) & $>82.0 (>69.0)$ & 2000 & $-1.48^{+0.20}_{-0.39}$ ($-1.48^{+0.59}_{-0.48}$) & $>79.2 (>65.8)$ \\
			
			1800 & $-1.57^{+0.16}_{-0.36}$ ($ -1.57^{+0.54}_{-0.42}$) & $>82.3 (>69.3)$ & 1800 & $-1.47^{+0.29}_{-0.34}$ ($-1.47^{+0.57}_{-0.47}$) & $>79.4 (>66.1)$ \\
			
			1600 & $-1.53^{+0.18}_{-0.38}$ ($-1.53^{+0.56}_{-0.45}$) & $>80.3 (>67.3)$ & 1600 & $-1.46^{+0.30}_{-0.36}$ ($-1.46^{+0.60}_{-0.48}$) & $>78.8 (>64.9)$ \\
			
			1400 & $-1.46^{+0.20}_{-0.42}$ ($-1.46^{+0.59}_{-0.50}$) & $>78.0 (>64.6)$ & 1400 & $-1.39^{+0.33}_{-0.38}$ ($-1.39^{+0.61}_{-0.54}$) & $>75.1 (>62.8)$ \\
			
			1200 & $-1.45^{+0.21}_{-0.42}$ ($-1.45^{+0.61}_{-0.50}$) & $>78.0 (>64.4)$ & 1200 & $-1.41^{+0.32}_{-0.38}$ ($-1.41^{+0.62}_{-0.52}$) & $>76.5 (>62.9)$ \\
			
			1000 & $-1.38^{+0.24}_{-0.47}$ ($-1.38^{+0.65}_{-0.55}$) & $>75.0 (>61.2)$ & 1000 & $-1.38^{+0.33}_{-0.39}$ ($-1.38^{+0.65}_{-0.54}$) & $>75.7 (>61.5)$ \\
			
			800  & $-1.44^{+0.22}_{-0.45}$ ($-1.44^{+0.66}_{-0.53}$) & $>77.1 (>62.5)$ & 800  & $-1.47^{+0.30}_{-0.37}$ ($-1.47^{+0.62}_{-0.51}$) & $>78.8 (>65.0)$ \\
			\hline
			\boldmath{$\ell_{\rm{min}}$} \textbf{(with \texttt{lowE})} & \boldmath{$w$}  &   \boldmath{$H_0$}~\textbf{[km/s/Mpc]} & \boldmath{$\ell_{\rm{min}}$} \textbf{(without \texttt{lowE})} & \boldmath{$w$}  &   \boldmath{$H_0$}~\textbf{[km/s/Mpc]}  \\
			\hline
			\hline
			30  & $-1.55^{+0.17}_{-0.39}$ ($-1.55^{+0.57}_{-0.45}$) & $>80.6 (>67.3)$ & 30  & $-1.28^{+0.32}_{-0.52}$ ($-1.28^{+0.68}_{-0.63}$) & $78^{+20}_{-9} (>57.4)$  \\
			
			200 & $-1.55^{+0.17}_{-0.38}$ ($-1.55^{+0.57}_{-0.44}$) & $>81.0 (>67.5)$ & 200 & $-1.02^{+0.39}_{-0.63}$ ($-1.02^{+0.78}_{-0.82}$) & $70^{+20}_{-20} (>48.4)$ \\
			
			400 & $-1.56^{+0.16}_{-0.36}$ ($-1.55^{+0.55}_{-0.43}$) & $>81.9 (>68.3)$ & 400 & $-1.08^{+0.39}_{-0.64}$ ($-1.08^{+0.80}_{-0.75}$) & $73^{+20}_{-10} (>49.6)$ \\
			
			600 & $-1.56^{+0.18}_{-0.38}$ ($-1.56^{+0.59}_{-0.46}$) & $>81.4 (>67.4)$ & 600 & $-1.07^{+0.41}_{-0.64}$ ($-1.07^{+0.80}_{-0.76}$) & $72^{+20}_{-10} (>49.7)$ \\
			
			800 & $-1.45^{+0.31}_{-0.53}$ ($-1.45^{+0.77}_{-0.67}$) & $>72.9 (>58.4)$ & 800 & $-1.07\pm 0.54$         ($-1.07^{+0.88}_{-0.91}$) & $70^{+20}_{-20} (>47.3)$ \\
			
			\hline\hline
	\end{tabular}}
	\caption{Constraints on the DE EoS $w$ at 68\% (95\%) CL, together with the corresponding 68\% (95\%) CL lower limits or constraints on the Hubble parameter $H_0$. The results are obtained by splitting the complete \textit{Planck} dataset into different subsets containing various multipole bins within the multipole range $\ell \in [\ell_{\min}, \ell_{\max}]$. When reducing $\ell_{\max}$, we fix $\ell_{\min}=2$. Conversely, when cutting low multipoles by increasing $\ell_{\min}$, we keep $\ell_{\max}=2500$ fixed. We repeat the same analysis, including and excluding the \texttt{lowE} likelihood for the EE spectrum at $2\le\ell\le30$.
	} 
	\label{table:values}
\end{table*}

\section{Results and discussion}
\label{sec:results}

\subsection{CMB analysis}
\label{subsec:cmb}

\begin{figure}[!ht]
	\centering
	\includegraphics[width=0.75\textwidth]{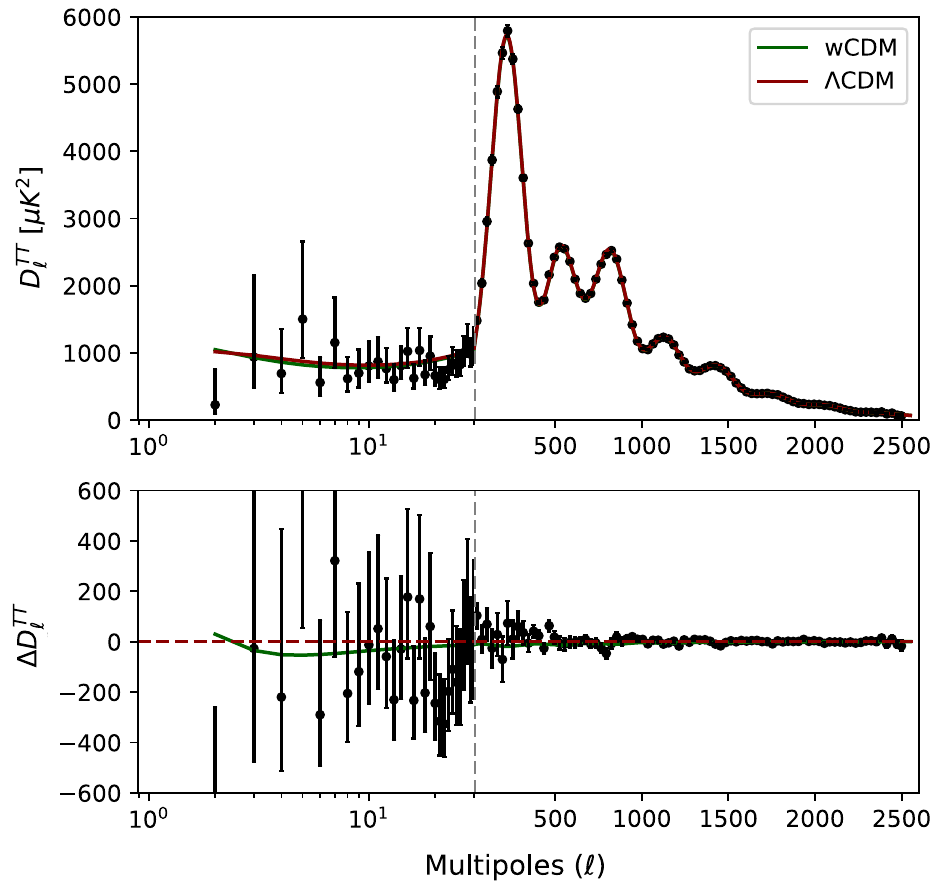}
	\caption{\textit{Upper panel}: Comparison between the $\Lambda$CDM (red curve) and $w$CDM (green curve) theoretical predictions for the temperature anisotropy power spectrum, fixing all cosmological parameters to their respective best-fit values based on the \textit{Planck} TTTEEE+\texttt{lowE} likelihoods. The error bars associated with the datapoints represent $\pm 1 \sigma$ uncertainties. \textit{Lower panel}: relative deviation between the $w$CDM and $\Lambda$CDM theoretical predictions. Note that at $\ell = 30$, for both panels the $x$-axis switches from logarithmic to linear scale.}
	\label{fig:cmb}
\end{figure}

We begin by examining the constraints derived exclusively using \textit{Planck} CMB data. All the results presented in this subsection are summarized in \autoref{table:values}.

First and foremost, it is important to highlight that when considering the full CMB dataset, we can find $68$\% and $95$\% confidence level (CL) constraint on the DE EoS of $w=-1.57^{+0.16}_{-0.36}$ and $w=-1.57^{+0.50}_{-0.40}$ respectively, together with a $95$\%~CL lower limit on the present-day expansion rate of the Universe, $H_0>69.3\,{\rm km}/{\rm s}/{\rm Mpc}$. As a result, we confirm consistent findings documented in the literature~\cite{Planck:2018vyg}, 
indicating a very mild preference in favor of a phantom EoS at a statistical level slightly exceeding two standard deviations.

The main purpose of this subsection is then to try to better comprehend the nature of this preference in \textit{Planck} data and evaluate its resilience. To take a first step forward in this direction, in \autoref{fig:cmb} we present a comparison between the best-fit to the \textit{Planck} temperature anisotropy power spectra obtained within the $w$CDM model (green curve) and the standard $\Lambda$CDM model (red curve). As one clearly sees from the Figure, both models appear nearly indistinguishable, which suggests from the outset that the degeneracy between $w$ and $H_0$ plays an important role in the interpretation of this result. In fact, parameters such as $\Omega_m$, $H_0$, $w$, and potentially others (such as the curvature parameter $\Omega_K$, not varied here), are affected by the geometrical degeneracy, 
~\cite{Bond:1997wr,Zaldarriaga:1997ch,Efstathiou:1998xx}. In this context, the effects on the spectrum of temperature anisotropies caused by highly negative values of $w$ can always be compensated by the opposite effect induced by an increase in $H_0$, leading to a significant degeneracy between these two parameters. This degeneracy, itself connected to the geometrical degeneracy, is the ultimate reason why $H_0$ remains unconstrained within $w$CDM.

However, upon closer inspection of \autoref{fig:cmb}, one can also note discernible differences between the two models at large angular scales (low multipoles), where the best-fit $w$CDM temperature anisotropy power spectrum exhibits a deficit of power at $\ell<30$ compared to the baseline $\Lambda$CDM case. Intriguingly, this goes in the direction of the long-standing and well-documented lack of power anomaly on large angular scales~\cite{Schwarz:2015cma}. At the lowest multipoles for the temperature spectrum, the amplitude of so-called Integrated Sachs-Wolfe (ISW) plateau is primarily controlled by the amplitude of the primordial scalar power spectrum $A_s$, the value of the spectral index $n_s$, and contributions resulting from the late-time ISW effect, which is instead sensitive to the DE dynamics. While the inferred values of both $A_s$ and $n_s$ remain nearly identical when moving from $\Lambda$CDM to $w$CDM, introducing a different DE EoS alters the late-time evolution of the Universe and thereby the decay of gravitational potentials which source the late ISW effect, in turn potentially altering the latter. We have explicitly verified that the observed differences between $\Lambda$CDM and $w$CDM in the fit at large angular scales are entirely due to the late ISW effect: we have switched off the ISW contribution to temperature anisotropies in \texttt{CAMB}, re-plotted \autoref{fig:cmb}, and noted that the two models become indistinguishable also at $\ell\lesssim 30$. This confirms our physical interpretation of the differences in fit between $\Lambda$CDM and $w$CDM at large angular scales being entirely attributable to the late ISW effect, which in the case of phantom DE is suppressed, as phantom DE opposes the decay of gravitational potentials, hence decreasing the late ISW source term.\footnote{This conclusion could be further strenghtened by explicitly varying a late ISW ``fudge factor'' amplitude, analogously to what was done in similar contexts in Refs.~\cite{Hou:2011ec,Cabass:2015xfa,Shajib:2016bes,Kable:2020hcw,Vagnozzi:2021gjh,Ruiz-Granda:2022bcn,Wang:2024kpu}.} Therefore, while the geometrical degeneracy between $H_0$ and $w$ appears to play a significant role in our inference of $w$, we cannot overlook the possibility that physical differences arising from a different EoS and potentially related to large-angle CMB anomalies may underlie (part of) this preference for $w<-1$. 

To continue our investigation, we have decided to maintain an agnostic approach and investigate how (if at all) our results are altered upon splitting the complete dataset into different subsets containing various multipole bins, and gradually subtracting information. Specifically, we consider CMB temperature and polarization anisotropy power spectra measured by \textit{Planck} within the multipole range $\ell \in [\ell_{\min}, \ell_{\max}]$, thus introducing two multipole cut-offs: $\ell_{\min}$, representing the minimum multipole considered in our analysis, and $\ell_{\max}$, which representing the maximum multipole analyzed.

\begin{figure}[ht!]
	\begin{center}
	\includegraphics[width=0.7\textwidth]{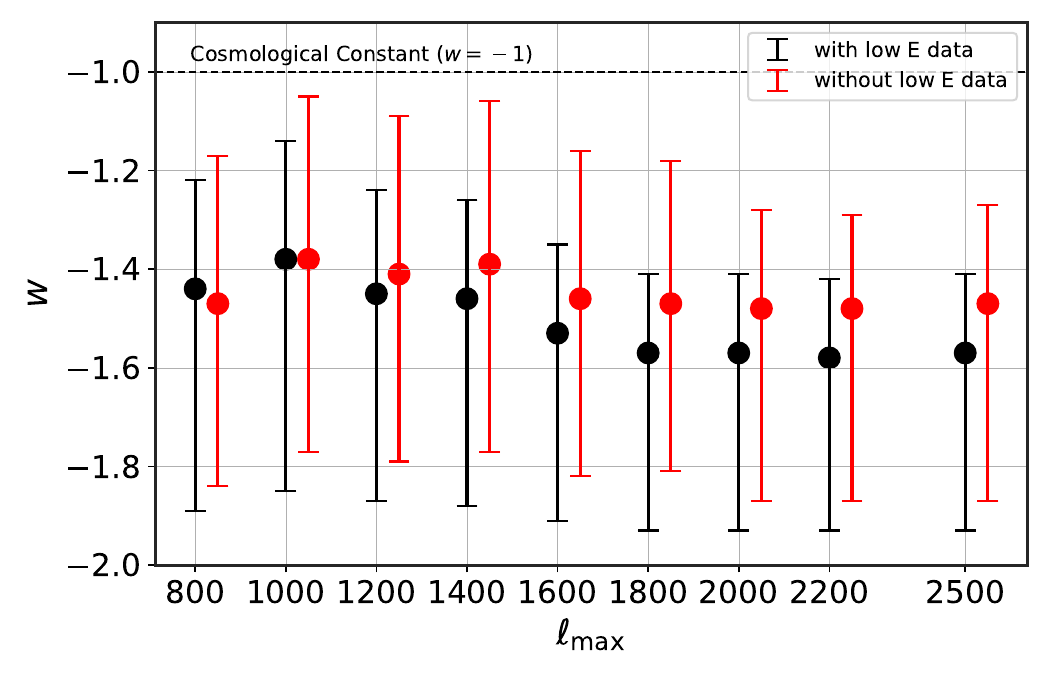}
	\caption{68\%~CL constraints on the dark energy equation of state $w$ obtained by fixing $\ell_{\rm{min}}=2$ and cutting information at high multipoles by considering different values of $\ell_{\rm{max}}$ ranging from 700 up to 2500 (which corresponds to the full datasets) The black (red) points represent the results obtained when including (excluding) the \texttt{lowE} likelihood for the EE polarization spectrum at $2 \leq \ell \leq 30$. An artificial small shift in the $x$-axis between red and black points is introduced solely for visualization purposes to prevent overlap.}
	\label{fig:w_lmax}
	\end{center}
\end{figure}

We begin by setting $\ell_{\min}=2$, thus including both the temperature (\texttt{lowT}) and polarization (\texttt{lowE}) \textit{Planck} likelihoods at $\ell < 30$. On the other hand, we truncate the spectra at high multipoles, considering progressively smaller values of $\ell_{\max}$, starting from $\ell_{\max}=2200$ and reducing it by 200 at each step until reaching the minimum value $\ell_{\max}=800$. The results presented in \autoref{table:values} and summarized in \autoref{fig:w_lmax} (focus for the moment only on the black points) demonstrate that by cutting off information from high multipoles, the preference for phantom DE is gradually diminished, but remains well above the one standard deviation level in all cases. This gradual reduction in the preference for $w<-1$ can be attributed to both the gradual increase in uncertainties (which is inevitable when considering only a subset of the full dataset), as well as a slight shift in the mean value of $w$ towards the cosmological constant value $w=-1$. Therefore, this test reveals the somewhat unexpected importance of precise measurements of the damping tail in order to accurately determine the DE EoS,\footnote{This conclusion is further supported by recent \textit{Planck}-independent CMB measurements released by ground-based telescopes, such as the Atacama Cosmology Telescope (ACT)~\cite{ACT:2020frw,ACT:2020gnv}. This experiment covers multipoles $\ell \gtrsim 650$ for temperature anisotropies and $\ell \gtrsim 350$ for polarization, thus collecting highly precise data in the damping tail, but exhibiting a lack of datapoints at low multipoles, particularly in the ISW plateau and around the first two acoustic peaks. Nevertheless, despite these limitations, it is still possible to derive a constraint on the DE EoS $w=-1.18^{+0.40}_{-0.55}$ from ACT data, see e.g.\ Tab.~2 of Ref.~\cite{DiValentino:2022rdg} or Fig.~1 of Ref.~\cite{Giare:2023xoc}.} and supports our initial caution with regards to attributing the \textit{Planck} preference for $w<-1$ solely to effects arising from the geometrical degeneracy.

In the same spirit, we now fix $\ell_{\max}=2500$ and progressively reduce the cut off at low multipoles. Starting with $\ell_{\rm{min}}=30$, we exclude ranges of multipoles in bins of 200 until we reach the extreme value of $\ell_{\min}=800$. This gradual exclusion of information at low multipoles eventually leads to the removal of data covering the first two acoustic peaks, thereby somewhat reducing the impact of the geometrical degeneracy. However, at this point it is worth noting that two different \textit{Planck} likelihoods cover information at multipoles $2 \leq \ell \leq 30$, namely the \texttt{lowT} likelihood for TT power spectrum measurements and the \texttt{Simall} \texttt{lowE}-likelihood for EE power spectrum measurements. As is well-known, measurements of E-mode polarization at large angular scales are crucial in order to infer the optical depth to reionization\footnote{Despite this conclusion remaining essentially true, some of us in Ref.~\cite{Giare:2023ejv} pointed out that it is indeed possible to derive constraints on $\tau$ independent of large-scale E-modes polarization with a relative precision comparable to the results obtained by \texttt{lowE}. As argued in the same work, temperature and polarization CMB data at $\ell>30$ and local Universe probes seem to prefer slightly higher values of $\tau$, possibly contributing to mitigating the anomalies observed in the Planck data.}, $\tau$~\cite{Lattanzi:2016dzq,Pagano:2019tci,Natale:2020owc,deBelsunce:2021mec,Wolz:2023gql}. This is due to the distinctive polarization bump produced by the epoch of reionization, which clearly manifests itself in the EE spectrum. Therefore, when it comes to reducing information at low multipoles, a choice needs to be made regarding whether or not to include the \texttt{lowE} likelihood, which has an important impact on constraints on $\tau$. We adopt a conservative approach and follow both paths.

\begin{figure}[ht!]
	\begin{center}
	\includegraphics[width=0.7\textwidth]{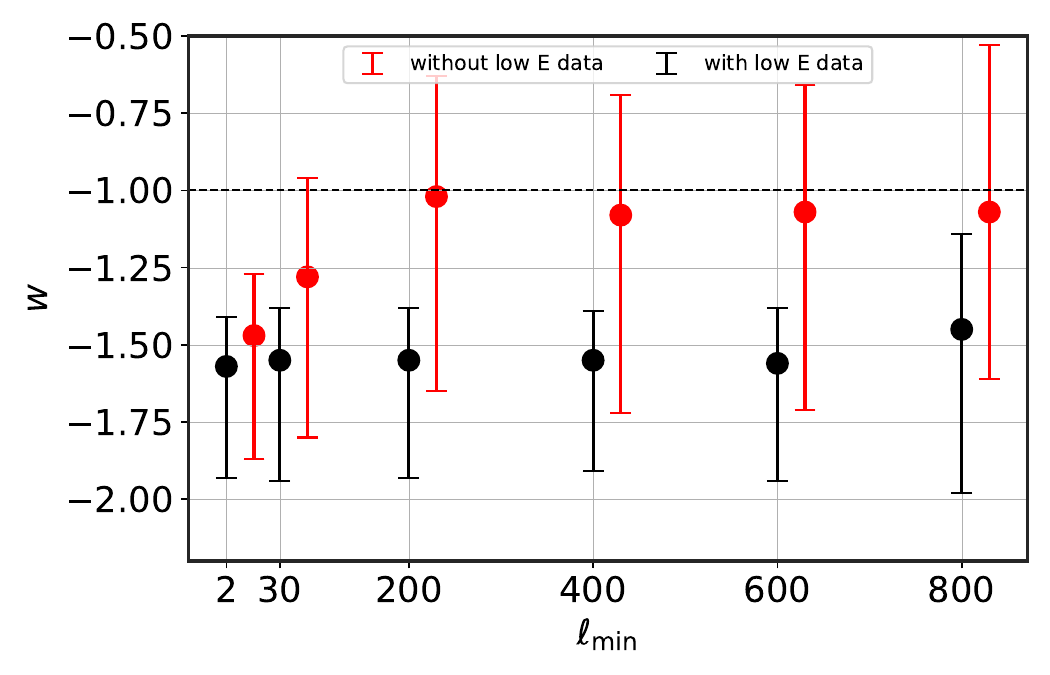}
	\caption{68\%~CL constraints on the dark energy equation of state $w$ obtained by fixing $\ell_{\rm{max}}=2500$ and cutting information at low multipoles by considering different values of $\ell_{\rm{min}}$ ranging from 2 (which corresponds to the full datasets) up to 800. The black (red) points represent the results obtained when including (excluding) the \texttt{lowE} likelihood for the EE polarization spectrum at $2 \leq \ell \leq 30$. An artificial small shift in the $x$-axis between red and black points is introduced solely for visualization purposes to prevent overlap.}
	\label{fig:w_lmin}
	\end{center}
\end{figure}

In the first case, we consistently include the \texttt{lowE} likelihood in the analysis to ensure that $\tau$ is well constrained by the data. Consequently, in this scenario, we consider EE data in the range $\ell \in [2,30]$ for low multipoles and $\ell \in[\ell_{\min},2500]$ for higher multipoles. Conversely, TT and TE spectra are always considered in the range $\ell \in [\ell_{\min},2500]$, meaning that the \texttt{lowT} likelihood is never included. By cutting off the information at low multipoles, we reach conclusions similar to those mentioned earlier. Specifically, the preference for a phantom EoS is gradually reduced, but always remains significantly above one standard deviation, as can be seen in \autoref{fig:w_lmin} (black points). This reduction is once more to be primarily attributed to increased uncertainties, with no noticeable shift observed in the central value, except for the most extreme case when $\ell_{\rm{min}}=800$, in which case the shift in the central value is nevertheless small.

\begin{figure}
	\begin{center}
	\includegraphics[width=0.7\textwidth]{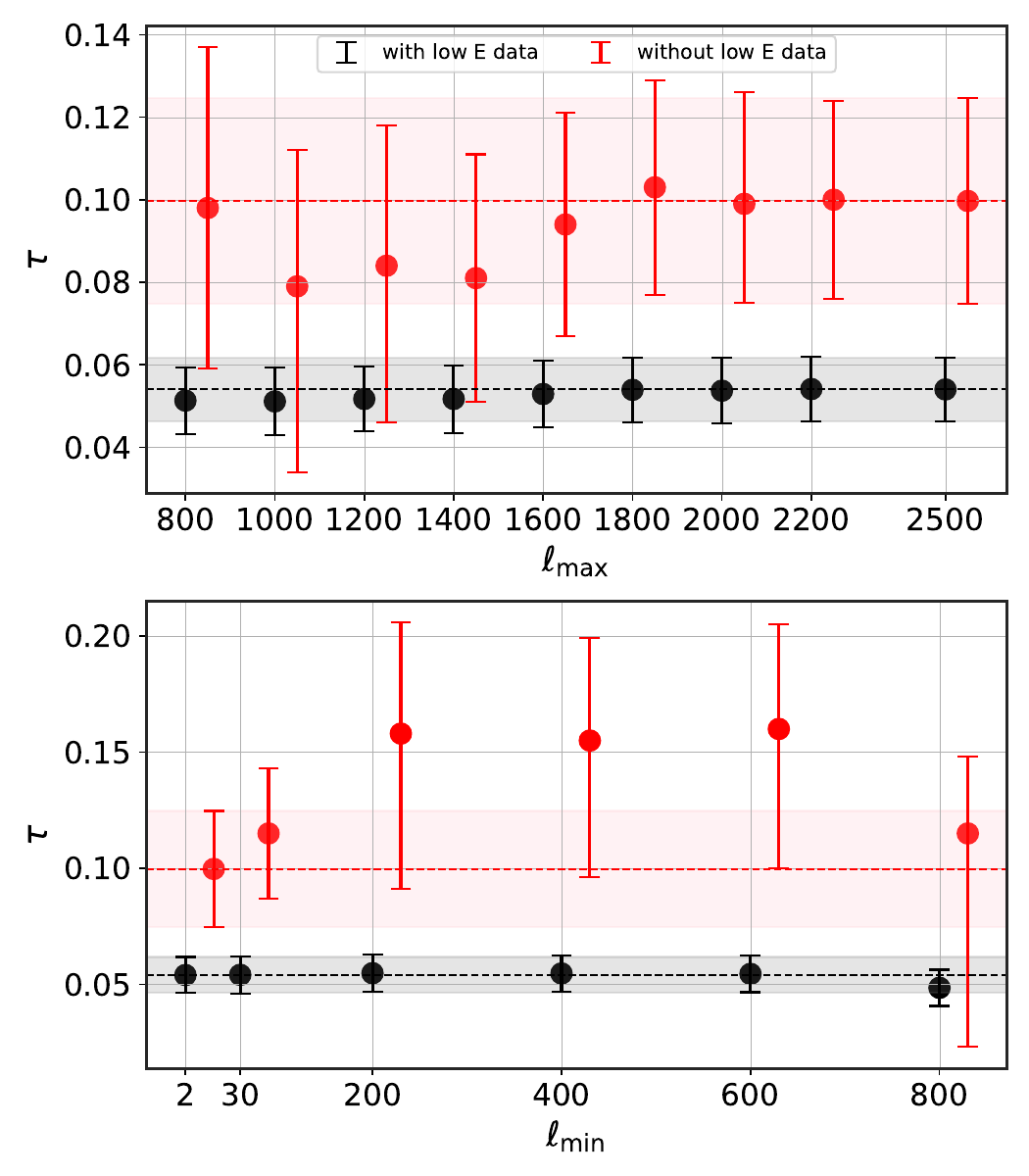}
	\caption{\textit{Upper panel}: 68\%~CL constraints on $\tau$ obtained by fixing $\ell_{\rm{min}}=2$ and cutting information at high multipoles by considering different values of $\ell_{\rm{max}}$. \textit{Lower panel}: As above, but fixing $\ell_{\rm{max}}=2500$ and cutting information at low multipoles by considering different values of $\ell_{\rm{min}}$. The black (red) points represent the results obtained when including (excluding) the \texttt{lowE} likelihood for the EE polarization spectrum at $2 \leq \ell \leq 30$. An artificial small shift in the $x$-axis between red and black points is introduced solely for visualization purposes to avoid overlap.}
	\label{fig:tau}
	\end{center}
\end{figure}

In the second case, we also exclude the \texttt{lowE} likelihood, resulting in a loss of constraining power for what concerns $\tau$, which now takes very large and unrealistic values $0.10 \lesssim \tau \lesssim 0.15$, as evident from \autoref{fig:tau} (lower panel). Interestingly, in this scenario, we observe a noticeable shift of $w$ towards the cosmological constant value $w=-1$, with no preference for a phantom EoS, as shown in \autoref{fig:w_lmin} (red points). This suggests that the exclusion of the \texttt{lowE} likelihood and, in general, relaxing constraints on the reionization epoch, may greatly influence the \textit{Planck} preference for phantom DE models, at the expense of unrealistically high values of $\tau$. This is due to the positive correlation between $\tau$ and the DE EoS: higher values of the former act to push $w$ towards the cosmological constant value.

Motivated by this intriguing finding, we repeat the previous analysis where we changed $\ell_{\max}$, this time excluding the \texttt{lowE} likelihood. Surprisingly, we observe different outcomes when cutting off high multipoles. The preference for phantom DE remains at above one standard deviation, as shown in \autoref{fig:w_lmax} (red points).

Overall, we conclude that constraints on the DE EoS from CMB data are influenced by a combination of effects originating from both low and high multipoles. These include the geometrical degeneracy (and thereby $H_0$), the late ISW effect and potentially low-$\ell$ anomalies, the lensing anomaly (see discussion in Appendix~\ref{sec:lensing}), high-$\ell$ measurements of the damping tail, and constraints on $\tau$ mainly coming from low-$\ell$ E-mode polarization measurements. Our conclusions are therefore twofold. Firstly, we have found it hard to isolate exactly where in the \textit{Planck} data the preference for phantom DE is arising from, but we have found that the latter is relatively resilient. Finally, and perhaps most importantly, given the role of the geometrical degeneracy, it is essential to include additional late-time cosmological probes to reach reliable conclusions concerning the apparent preference for phantom DE in \textit{Planck} data~\cite{Planck:2018vyg}.

\subsection{Joint CMB and late-time probes analysis}
\label{subsec:cmblateprobes}

\begin{table*}[!ht]
	\footnotesize
	\centering 
	\renewcommand{\arraystretch}{1.5}
	\resizebox{0.75 \textwidth}{!}{
		\begin{tabular}{ccc} 
			\textbf{Dataset combination} & \boldmath{$w$} & \boldmath{$H_0\,[{\rm km}/{\rm s}/{\rm Mpc}]$} \\
			\hline\hline
			CMB & $-1.57^{+0.16}_{-0.36}$ ($-1.57^{+0.53}_{-0.42}$) & $>82.4 \quad (>69.3)$ \\
			
			CMB+BAO & $-1.039\pm 0.059$  ($-1.04^{+0.11}_{-0.12}$)  & $68.6\pm 1.5 \quad (68.6^{+3.1}_{-2.8})$ \\
			
			CMB+SN & $-0.976\pm 0.029$   ($-0.976^{+0.055}_{-0.056}$)  &   $66.54\pm 0.81 \quad (66.5^{+1.6}_{-1.6})$ \\
			
			CMB+CC & $-1.21\pm 0.19$ ($-1.21^{+0.35}_{-0.38}$)  & $73.8\pm 5.9 \quad (74^{+10}_{-10})$ \\
			
			CMB+BAO+SN & $-0.985\pm 0.029$ ($-0.985^{+0.054}_{-0.062}$) & $67.40^{+0.90}_{-1.1} \quad  (67.4^{+2.0}_{-1.8})$ \\
			
			\textbf{\underline{CMB+BAO+SN+CC}} & $\boldsymbol{-1.013^{+0.038}_{-0.043}}$  $\boldsymbol{(-1.013^{+0.095}_{-0.098}})$ & $\boldsymbol{68.6^{+1.7}_{-1.5} \quad  (68.6^{+3.0}_{-3.1})}$ \\
			
			CMB+BAO+SN+SH0ES & $-1.048\pm 0.027$ ($-1.048^{+0.049}_{-0.054}$) & $69.43\pm 0.66 \quad  (69.4^{+1.3}_{-1.3})$ \\
			
			CMB+FS & $-0.991\pm 0.043$ ($-0.991^{+0.077}_{-0.085}$) & $67.2\pm 1.1 \quad (67.2^{+2.3}_{-2.1})$ \\
			
			BAO+SN & $-0.95^{+0.14}_{-0.11}$ ($-0.95^{+0.22}_{-0.24}$) & $67.2^{+5.3}_{-6.1} \quad (67.2^{+7.1}_{-6.7})$ \\
			
			BAO+SN+CC & $-0.948^{+0.13}_{-0.10}$ ($-0.95^{+0.15}_{-0.21}$) & $67.0^{+3.9}_{-5.6} \quad (67^{+10}_{-8})$ \\
			
			BAO+CC & $-0.86\pm 0.12$ ($-0.86^{+0.23}_{-0.24}$) & $64.0\pm 3.3 \quad (64.0^{+6.6}_{-6.5})$ \\
			\hline\hline
	\end{tabular}}
	\caption{68\% (95\%)~CL constraints on the dark energy equation of state $w$, together with the corresponding 68\% (95\%)~CL lower limits or constraints on the Hubble parameter $H_0$ obtained by considering different combinations of late-time probes, either in combination with \textit{Planck} CMB data or alone. When CMB data is not included in the analysis, the following parameters are fixed to their \textit{Planck} TTTEEE + \texttt{lowE} best-fit values: $\Omega_b h^2 = 0.022377$, $\log(10^{10}A_s)=3.0447$, $n_s=0.9659$, and $\tau=0.0543$.}
	\label{table:values2}
\end{table*}

Based on the previous analysis of CMB observations alone, this subsection focuses on studying late-time cosmological probes, both alone and in combination with CMB data, in order to assess the robustness of the apparent preference for phantom DE from \textit{Planck} data alone. Our aim here is to investigate whether such a preference remains when the geometrical degeneracy involving $w$ is broken by including additional late-time observations. We consider a wide range of late-time probes, as discussed earlier in Sec.~\ref{sec:data}, and our results are summarized in \autoref{table:values2}. We proceed by scrutinizing the impact of adding these datasets (one at a time) to CMB data. Subsequently, we discuss the results obtained by combining multiple late-time datasets, before finally obtaining constraints using only late-time probes.

\subsubsection{Baryon Acoustic Oscillations (BAO)}
\label{subsubsec:bao}

\begin{figure}[!t]
	\begin{center}
		\centering
		\includegraphics[width=0.8\textwidth]{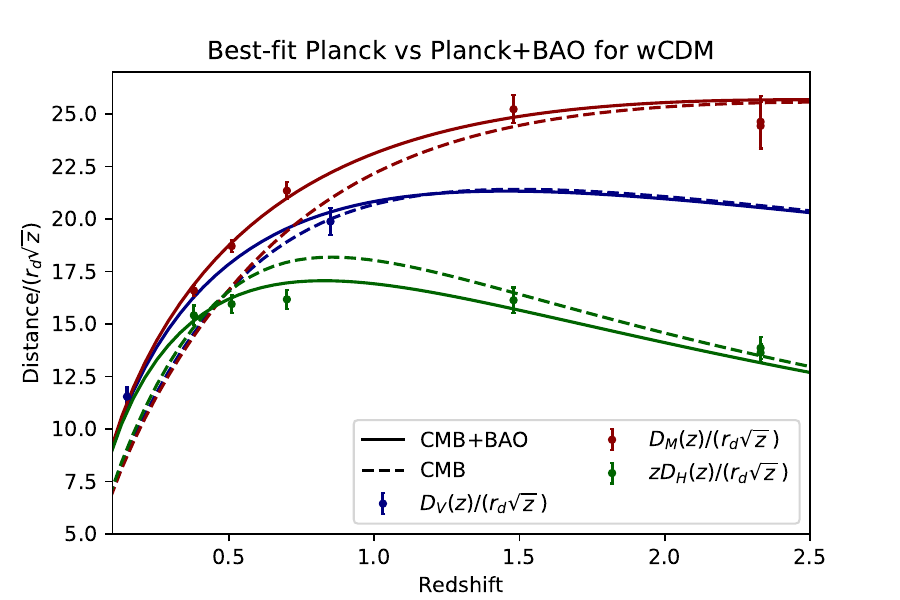}
		\caption{Best-fit predictions for (rescaled) distance-redshift relations from a $w$CDM fit to \textit{Planck} CMB data alone (dashed curves) and the CMB+BAO dataset (solid curves). These predictions are presented for the three different types of distances probed by BAO measurements (rescaled as per the $y$ label), each indicated by the colors reported in the legend. The error bars represent $\pm 1 \sigma$ uncertainties.}
		\label{fig:baowcdmplanckbao}
	\end{center}
\end{figure}

\begin{figure}[!t]
	\centering
	\includegraphics[width=0.8\textwidth]{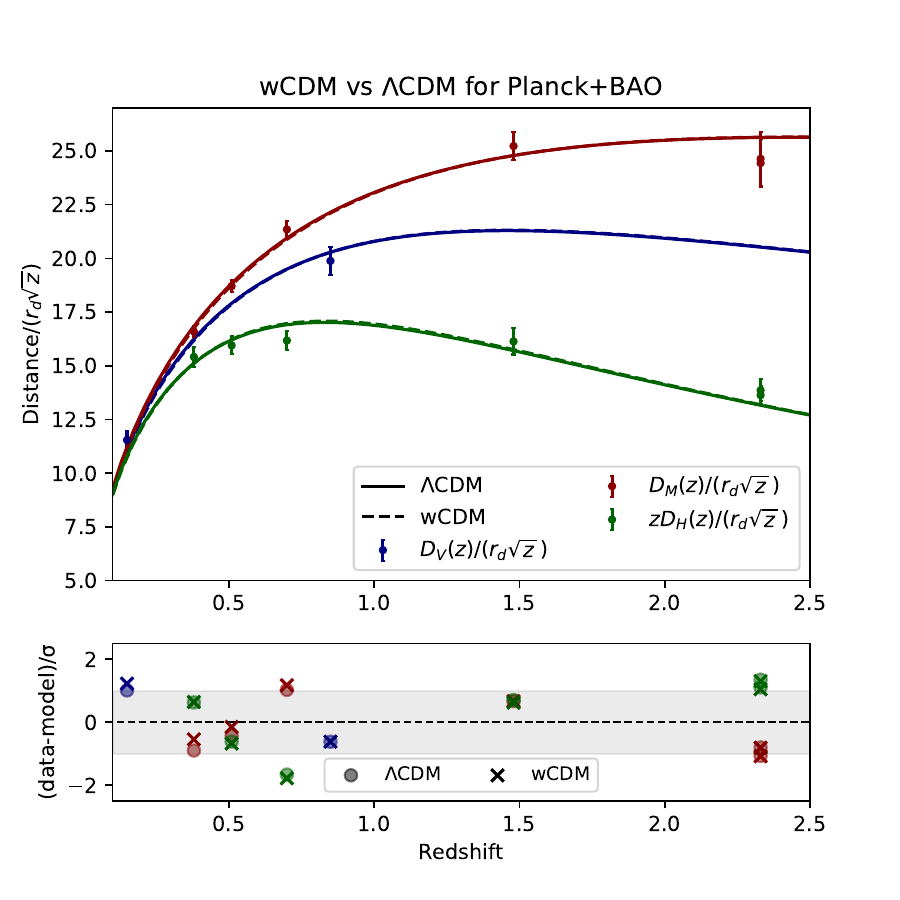}
	\caption{\textit{Upper panel}: Best-fit predictions for (rescaled) distance-redshift relations from $\Lambda$CDM (solid curves) and $w$CDM (dashed curves) to the CMB+BAO dataset. These predictions are presented for the three different types of distances probed by BAO measurements, each indicated by the colors reported in the legend. The error bars represent $\pm 1 \sigma$ uncertainties. \textit{Lower panel}: Difference between the model prediction and datapoint for each BAO measurement, normalized by the observational uncertainties. The $\Lambda$CDM predictions are represented by `o'-shaped points, while the $w$CDM predictions are represented by `x'-shaped points.}
	\label{fig:baolcdmwcdmplanckbao}
\end{figure}

We now consider the CMB+BAO dataset combination. The parameter inference performed in this subsection is similar to that of Ref.~\cite{eBOSS:2020yzd}, with the main distinction being the sampler (\texttt{Cobaya} instead of \texttt{CosmoMC}). We infer $w=-1.039 \pm 0.059$, consistent with the cosmological constant value $w=-1$ within one standard deviation. From the same dataset, we infer $H_0=(68.6 \pm 1.5)\,{\rm km}/{\rm s}/{\rm Mpc}$. 
As we saw earlier in \autoref{fig:cmb}, the difference between the best-fit $\Lambda$CDM and $w$CDM CMB temperature power spectra were negligible (if not for small changes due to the late ISW effect, anyhow completely swamped by cosmic variance): this is a direct reflection of the geometrical degeneracy. On the other hand, late-time distance and expansion rate measurements are very effective at breaking this degeneracy. To show this, in \autoref{fig:baowcdmplanckbao} we plot the best-fit predictions for (rescaled) distance-redshift relations from a $w$CDM fit to \textit{Planck} CMB data alone (dashed curves), and the CMB+BAO dataset (solid curves), for the three different types of distances probed by BAO measurements. These distances are: the comoving angular diameter distance $D_M(z)=D_A(z)(1+z)$, which measures the spatial distance between two objects in the direction perpendicular to the line-of-sight; the line-of-sight distance $D_H(z)=c/H(z)$, which measures the distance along the line-of-sight between an observer and an object; and the volume-averaged distance $D_V(z)=[zD_H(z)D_M^2(z)]^{1/3}$, which is the quantity to which isotropic BAO measurements are sensitive. We clearly see that the predicted distances vary enormously within the two models, well outside of the uncertainty range of BAO measurements. Therefore, it should not come as a surprise that the CMB+BAO dataset combination rules out the deep phantom model preferred by CMB data alone at high statistical significance.

Intriguingly, although the value of the DE EoS inferred from the CMB+BAO combination is consistent with the cosmological constant value within one standard deviation, the central value is slightly in the phantom region, $w \sim -1.04$, in the same ballpark discussed towards the end of Sec.~\ref{sec:deeos}. To further understand this constraint, in the upper panel of \autoref{fig:baolcdmwcdmplanckbao} we plot the best-fit predictions for (rescaled) distance-redshift relations from $w$CDM (dashed curves) and $\Lambda$CDM fits (solid curves) to the CMB+BAO dataset. The difference in goodness-of-fits between the two models is hard to distinguish by the naked eye, even when considering the uncertainty-normalized residuals (data minus model) in the lower panel of the same figure.

By inspecting the lower panel of \autoref{fig:baolcdmwcdmplanckbao}, we see that there is no single datapoint for which the $w$CDM fit is noticeably better than the $\Lambda$CDM one, and which could therefore drive the central value of $w$ towards the very mild phantom region ($w \sim -1.04$). The only datapoint which could be marginally interesting this sense is the left-most maroon ($D_M/r_d$) datapoint, corresponding to the lowest redshift ($0.2<z<0.5$) BOSS galaxy sample~\cite{eBOSS:2020yzd}. It has long been known that the three maroon points, corresponding to two BOSS galaxy and one eBOSS galaxy samples (in the redshift ranges $0.2<z<0.5$, $0.4<z<0.6$, and $0.6<z<1.0$ respectively), exhibit a ``linear'' trend which appears to make them slightly deviate from the $\Lambda$CDM predictions (precisely in the direction predicted by phantom DE), albeit at a very weak ($\gtrsim 1\sigma$) statistical significance: this trend can be seen quite clearly when considering the three maroon points in \autoref{fig:baolcdmwcdmplanckbao} (both the upper and lower panels). One could argue that $w$CDM fits these three points \textit{ever slightly better} than $\Lambda$CDM, and therefore that it is these points that drive the central value of $w$ towards the very mild phantom region. However, given the extremely weak statistical significance of the latter ($\lesssim 0.7\sigma$), we choose to not delve on this aspect further.

\subsubsection{Type Ia Supernovae (SN)}
\label{subsubsec:sneia}

\begin{figure}[!ht]
	\centering
	\includegraphics[width=0.75\textwidth]{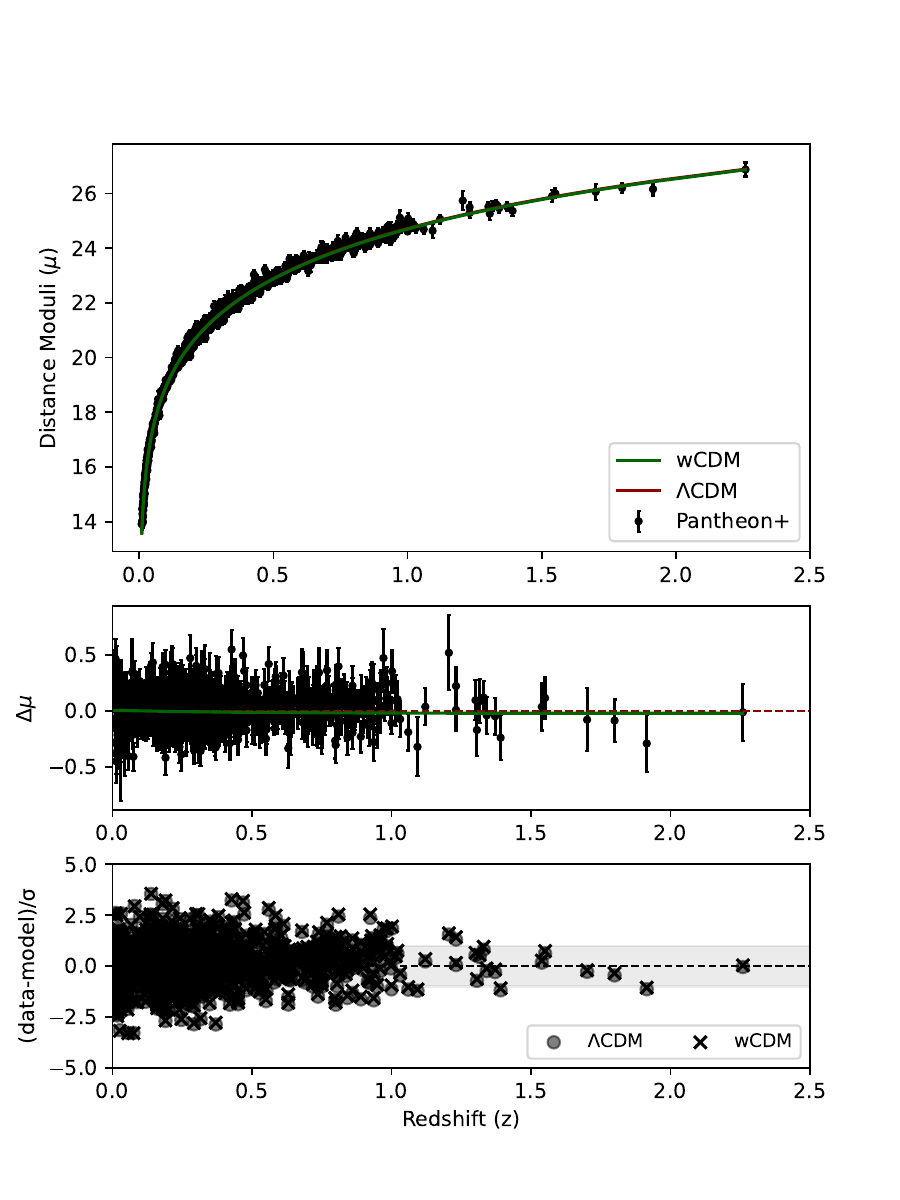}
	\caption{\textit{Upper panel}: Best-fit predictions for the distance-moduli from $\Lambda$CDM (red curve) and $w$CDM (green curve) fits to the the CMB+SN dataset. \textit{Middle panel}: Residuals for the two different models. \textit{Lower panel}: Difference between the model prediction and datapoint for each SN measurement, normalized by the observational uncertainties. The $\Lambda$CDM predictions are represented by `o'-shaped points, while the $w$CDM predictions are represented by `x'-shaped points.}
	\label{fig:sneia}
\end{figure}

We now consider the CMB+SN dataset combination. The parameter inference performed in this subsection, much like the previous case, is similar to that of Ref.~\cite{Brout:2022vxf}, with the only difference being the sampler used. As BAO measurements, distance moduli measurements (and therefore effectively uncalibrated luminosity distance measurements) are very efficient at breaking the geometrical degeneracy and thereby increase the fidelity of the inferred value of $w$. From this combination we infer $w=-0.976 \pm 0.029$, again consistent with the cosmological constant value within one standard deviation, although this time with central value slightly in the quintessence-like region, unlike what we obtained from the CMB+BAO combination. For what concerns the Hubble constant, we instead infer $H_0=(66.54 \pm 0.81)\,{\rm km}/{\rm s}/{\rm Mpc}$. We stress that the value of $H_0$ inferred is low as we have not used the SH0ES calibration. Had we used it (see later discussion in \autoref{subsubsec:joint}), we can expect the inferred value of $w$ to move towards the phantom regime, given the inverse correlation between $w$ and $H_0$, extensively documented in the literature~\cite{DiValentino:2016hlg,Zhao:2017cud,Vagnozzi:2018jhn,Vagnozzi:2019ezj,Banerjee:2020xcn}.

As in the previous subsection, in \autoref{fig:sneia} we plot the best-fit predictions for the distance moduli-redshift relation from $w$CDM (green curve) and $\Lambda$CDM (maroon curve) fits to the CMB+SN dataset. As in the CMB+BAO case, the difference in goodness-of-fits between the two models is virtually impossible to distinguish by the naked eye, even when plotting the residuals (in the middle and lower panels of the same figure, in one case uncertainty-normalized and in the other one not). In particular, there is no obvious set of datapoints/redshift range for which a quintessence-like DE component appears to fit the data significantly better than $\Lambda$CDM. Given the extremely weak statistical significance of the indication for $w \neq -1$ ($\lesssim 0.8\sigma$), we do not explore this aspect further.

\subsubsection{Cosmic Chronometers (CC)}
\label{subsubsec:cc}

\begin{figure}[!t]
	\centering
	\includegraphics[width=0.7\textwidth]{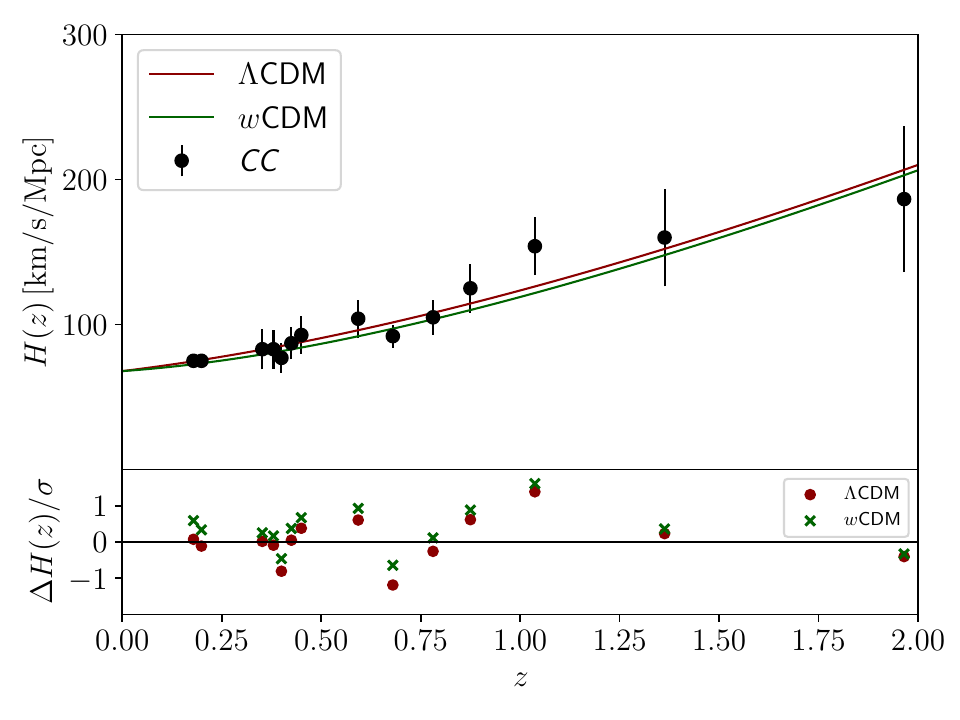}
	\caption{\textit{Upper panel}: Best-fit predictions for the redshift evolution of the Hubble parameter $H(z)$ for the $\Lambda$CDM (red curve) and $w$CDM (green curve) fits to the the CMB+CC dataset. \textit{Bottom panel}: Difference between the model prediction and the datapoint for each CC measurement, normalized by the observational uncertainties. The $\Lambda$CDM predictions are represented by `o'-shaped points, while the $w$CDM predictions are represented by `x'-shaped points.}
	\label{fig:cc}
\end{figure}

We now move on to consider the CMB+CC dataset combination. The parameter inference performed here is substantially different from that of Ref.~\cite{Vagnozzi:2020dfn}, which focused on the role of spatial curvature but also reported, in passing, $w$CDM-based constraints. The two key differences are that we only take a subset (15 points) of the 31 points used in Ref.~\cite{Vagnozzi:2020dfn}, i.e.\ only those with full covariance matrix available; and for analyzing these points we of course make use of the full non-diagonal covariance matrix, as opposed to the diagonal covariance matrix used in Ref.~\cite{Vagnozzi:2020dfn})

As previously with distance measurements, we note that expansion rate measurements are also extremely effective in breaking the geometrical degeneracy, as also stated in~\cite{Vagnozzi:2020dfn}, and improving the fidelity of the inferred value of the DE EoS. From the CMB+CC combination we infer $w=-1.21 \pm 0.19$, exhibiting a very weak ($\gtrsim 1.1\sigma$) indication for phantom DE, albeit overall broadly consistent with a cosmological constant in light of the relatively large uncertainties, despite the central value being quite deep in the phantom regime. The inferred Hubble constant is instead $H_0=(73.8 \pm 5.9)\,{\rm km}/{\rm s}/{\rm Mpc}$.

In \autoref{fig:cc} we repeat the same analysis of the residuals performed in the previous subsections, plotting the best-fit predictions for the expansion rate as a function of redshift for $w$CDM (blue curve) and $\Lambda$CDM (red curve) fits to the CMB+CC dataset. In this case, especially given the significantly smaller number of datapoints compared to the SN dataset, the difference between the $\Lambda$CDM and $w$CDM predictions is more easily discernible by eye. Nevertheless, even when considering the uncertainty-normalized residuals in the lower panel of \autoref{fig:cc}, it remains difficult to tell which datapoints/redshift range is driving the inference of $w$ towards the phantom region. The only datapoints which could be marginally interesting in this sense are the two measurements around $z=0.75$: specifically, these are two datapoints at $z=0.68$ and $z=0.78$ respectively, both compiled in Ref.~\cite{Moresco:2012by}. In particular, the $\Lambda$CDM prediction for the expansion rate at $z=0.68$ is $\sim 1.5\sigma$ off from the measured value, whereas the $w$CDM prediction is consistent within $\sim 0.5\sigma$. Nevertheless, given the very weak statistical significance of the indication for $w \neq -1$, we do not delve any further on the question of which parts of the data are driving the fit towards phantom DE.

\subsubsection{Full-shape galaxy power spectrum (FS)}
\label{subsubsec:fs}

The next dataset combination we consider is the CMB+FS one, which of course includes (at least in part) the geometrical information already contained in the CMB+BAO combination. From the CMB+FS combination we infer $w=-0.991 \pm 0.043$, in excellent agreement with the cosmological constant value $w=-1$, even at the level of central value. Moreover, we note that this inference is in very good agreement, within $\lesssim 0.7\sigma$, with the value obtained from the CMB+BAO combination. For the Hubble constant was instead obtained $H_0=(67.2 \pm 1.1)\,{\rm km}/{\rm s}/{\rm Mpc}$.

Two comments are in order regarding this dataset combination. Firstly, we choose not to perform an analysis of residuals analogous to the ones summarized in the previous subsections in \autoref{fig:baolcdmwcdmplanckbao}, \autoref{fig:sneia}, and \autoref{fig:cc}. The first reason is simply that the value of $w$ inferred from CMB+FS is in so good agreement with the cosmological constant value, that making such an analysis is somewhat unwarranted.

The second reason why we choose not to perform such an analysis is more subtle, and has to do with the treatment of nuisance parameters in the FS likelihood. In short, the EFTofLSS approach introduces a large number of nuisance parameters.\footnote{Most of these are basically coefficients of a series of operators composed of long-wavelength fields, whose functional form is fixed, and which model short-scale physics without the need for a detailed description of the latter but simply resort to symmetry arguments.} Most of these nuisance parameters, such as the third-order tidal bias term, the $\ell=0\,,2\,,4$ counterterms, the Fingers-of-God counterterm, and various (scale-independent and scale-dependent) shot noise parameters, are analytically marginalized over in the FS likelihood~\cite{Ivanov:2019pdj,Philcox:2020vvt,Chudaykin:2020aoj,Ivanov:2021fbu,Philcox:2021kcw,Cabass:2022wjy,Philcox:2022frc}. In order to make MCMC analyses feasible, the default version of the FS likelihood only explicitly varies a handful of nuisance parameters per galaxy sample, e.g.\ (for the power spectrum) the linear bias, quadratic bias, and quadratic tidal bias terms. If all the nuisance parameters were explicitly varied, for a power spectrum and bispectrum joint analysis the number of nuisance parameters would easily exceed 100, resulting in an extremely long convergence time for MCMC runs, which would become highly impractical. However, the fact that most nuisance parameters are marginalized over makes a ``fair'' comparison of residuals (data minus model) across different models impossible. One could envisage keeping the analytically marginalized EFTofLSS nuisance parameters fixed to a reference value when moving across two (or more) models. But this would not be representative of the true situation, as one expects nuisance parameters to shift and partially (re-)absorb the differences between theoretical predictions for the observables: this point was explicitly discussed in a number of recent works, e.g.\ Refs.~\cite{Ivanov:2020ril,Nunes:2022bhn,Reeves:2022aoi}. In light of these two reasons, we opted for not comparing the $\Lambda$CDM vs $w$CDM residuals for the CMB+FS combination.

The second comment pertains to comparison against earlier results obtained using \textit{similar} dataset combinations. A number of works have recently examined constraints on $w$ in light of FS measurements, with the specific constraints varying across different works (depending on the underlying assumptions), but all falling within the phantom regime. For the sake of concreteness, we quote the following recent results:
\begin{itemize}
	\item Ref.~\cite{DAmico:2020kxu}: $w=-1.05^{+0.06}_{-0.05}$ from a fit to BOSS FS and \textit{Pantheon} SNeIa data, alongside a BBN prior on the physical baryon density, within the $w$CDM model, using the \texttt{PyBird} code~\cite{DAmico:2020kxu}. Without the inclusion of SNeIa data the result moves slightly more into the phantom regime, $w=-1.10^{+0.14}_{-0.11}$.
	\item Ref.~\cite{Chudaykin:2020ghx}: $w=-1.03 \pm 0.05$ from a fit to BOSS FS and \textit{Pantheon} SNeIa data, alongside a BBN prior on the physical baryon density, within the $w$CDM model, using the \texttt{CLASS-PT} code~\cite{Chudaykin:2020aoj}. Without the inclusion of SNeIa data the result moves slightly more into the phantom regime, $w=-1.04^{+0.10}_{-0.08}$.
	\item Ref.~\cite{Brieden:2022lsd}: $w=-1.09 \pm 0.05$ from a fit to \textit{Planck} CMB data, as well as BOSS and eBOSS galaxy clustering data, and a BBN prior on the physical baryon density, within the $w$CDM model, adopting the ShapeFit one-parameter extension of the standard approach, as discussed in Refs.~\cite{Brieden:2021edu,Brieden:2022ieb}. When CMB data is not included, the inference changes to $w=-1.00^{+0.09}_{-0.07}$.
	\item Ref.~\cite{Carrilho:2022mon}: $w=-1.17 \pm 0.12$ from a fit to BOSS FS data alone within the $w$CDM model. However, this inference was argued to be affected by prior volume effects, as the best-fit $w=-1.09$ is slightly less phantom, although both values remain quite deep in the phantom regime.
	\item Ref.~\cite{Semenaite:2022unt}: $w=-1.07^{+0.06}_{-0.05}$ from a fit to \textit{Planck} CMB data, as well as BOSS and eBOSS galaxy clustering data within the $w$CDM model. When the curvature parameter is varied as well, $w$ moves deeper into the phantom regime, $w=-1.10^{+0.08}_{-0.07}$.
\end{itemize}
Other constraints on $w$ have been obtained from analyses of FS datasets in the literature, but here we choose to just report these five representative results by way of example.

It is interesting to note that virtually all of the above analyses inferred a value of $w$ within the phantom regime when using FS data. This is in contrast to our results, where we find a value of $w$ perfectly consistent with the cosmological constant, and a central value actually slightly within the quintessence-like regime. With the caveat that we cannot perform a residuals analysis for the reason discussed previously, here we simply note that a direct comparison to all the above results is in principle very tricky. For instance, none of the above analyses made use of bispectrum data (which, to the best of our knowledge, is being used for the first time here to constrain $w$). Furthermore, some of the above made different assumptions concerning the theoretical modelling (e.g.\ full EFTofLSS analysis vs ShapeFit three-parameter compression). Finally, different codes were used in some of the above analyses compared to ours. Therefore, we believe that a comparison to the above results, if at all, should be performed with considerable caution.

\subsubsection{Joint Analyses}
\label{subsubsec:joint}

Having examined various combinations of CMB data and other external datasets, one at a time, we now consider joint analyses of dataset combinations other than the ones studied earlier. We begin by considering the CMB+BAO+SN dataset combination (for which a similar analysis can be found in Ref.~\cite{Brout:2022vxf}), from which we infer $w=-0.985 \pm 0.029$, which agrees within $\approx 0.5\sigma$ with the cosmological constant value. If we additionally include the SH0ES calibration to the \textit{PantheonPlus} sample, i.e.\ considering the CMB+BAO+SN+SH0ES dataset combination (which can be compared to the analysis performed in Ref.~\cite{Brout:2022vxf}), we find $w=-1.048 \pm 0.027$. Unsurprisingly, $w$ moves into the phantom regime, again due to the inverse correlation between $w$ and $H_0$~\cite{DiValentino:2016hlg,Zhao:2017cud,Vagnozzi:2018jhn,Vagnozzi:2019ezj,Banerjee:2020xcn}, and easily understandable by the effects of $w$ and $H_0$ on the expansion rate and therefore cosmic distances.

We then consider the CMB+BAO+SN+CC dataset combination. Overall, given the effectiveness of this combination breaking the geometrical degeneracy, as well as the overall trustworthiness of the dataset (recall that we use only those CC datapoints for which the full covariance matrix, including the effects of systematics, is available), we select this dataset combination as being the one from which we quote our final consensus results. In particular, we infer $w=-1.013^{+0.038}_{-0.043}$, in excellent agreement (within $\approx 0.3\sigma$) with the cosmological constant value, even though the central value still falls within the phantom regime, and we observe a slight uptick in the error-bars, which could suggest some minor challenges when attempting to accommodate all the datasets within the same fit simultaneously. From the same dataset combination, we also infer a value of the Hubble constant $H_0=68.6^{+1.7}_{-1.5}\,{\rm km}/{\rm s}/{\rm Mpc}$, in good agreement with the model-dependent $\Lambda$CDM-based determination.

Finally, for completeness, we also consider three CMB-free dataset combination, from which constraints on $w$ are therefore obtained purely based on geometrical information. For these three combinations, we fix the values of $\Omega_bh^2$, $\log(10^{10}A_s)$, $n_s$, and $\tau$ to their \textit{Planck} TTTEEE+\texttt{lowE} best-fit values, i.e.\ $\Omega_b h^2 = 0.022377$, $\log(10^{10}A_s)=3.0447$, $n_s=0.9659$, and $\tau=0.0543$. The first combination we consider is the BAO+SN one (a similar analysis was conducted in Ref.~\cite{Staicova:2021ntm} but with the older \textit{Pantheon} dataset, for a more recent one making use of the \textit{PantheonPlus} dataset the reader can refer to Ref.~\cite{Cogato:2023atm}), from which we infer $w=-0.95^{+0.14}_{-0.11}$, in excellent agreement with the cosmological constant value. Similar considerations hold for the BAO+SN+CC dataset combination (for a similar analysis albeit with the older \textit{Pantheon} dataset please refer to Refs.~\cite{Benisty:2020otr,LozanoTorres:2023jwh}, and to Ref.~\cite{Cogato:2023atm} for one with \textit{PantheonPlus}), from which we infer $w=-0.948^{+0.13}_{-0.10}$. Finally, for completeness we also consider the BAO+CC combination (for a similar analysis please refer to Ref.~\cite{Cogato:2023atm}), finding $w=-0.86\pm 0.12$. Here the agreement with the cosmological constant value is still very good (within better than $1.5\sigma$), although less so than for the other two cases. The final take-away point we note is that for all three CMB-free dataset combinations, the central value of $w$ lies within the quintessence-like regime: we take this as a good indication that the overall very weak indication for phantom DE found in all other combinations was mostly driven by the \textit{Planck} CMB dataset, for the reasons discussed in Sec.~\ref{subsec:cmb} and Appendix~\ref{sec:lensing}, including the geometrical degeneracy (and thereby $H_0$), the late ISW effect and potentially low-$\ell$ anomalies, the lensing anomaly, high-$\ell$ measurements of the damping tail, and constraints on $\tau$ mainly coming from low-$\ell$ E-mode polarization measurements. However, given the extremely weak preference for quintessence-like DE in the CMB-free cases, we have been unable to pinpoint the origin of such a feature (for the same reason that we have found it very challenging to identify the origin of the stronger preference for phantom DE in the cases involving the CMB, even though the significance was indeed stronger).

\section{Conclusions}
\label{sec:conclusions}

\begin{figure*}[!t]
\centering
\includegraphics[width=0.8\textwidth]{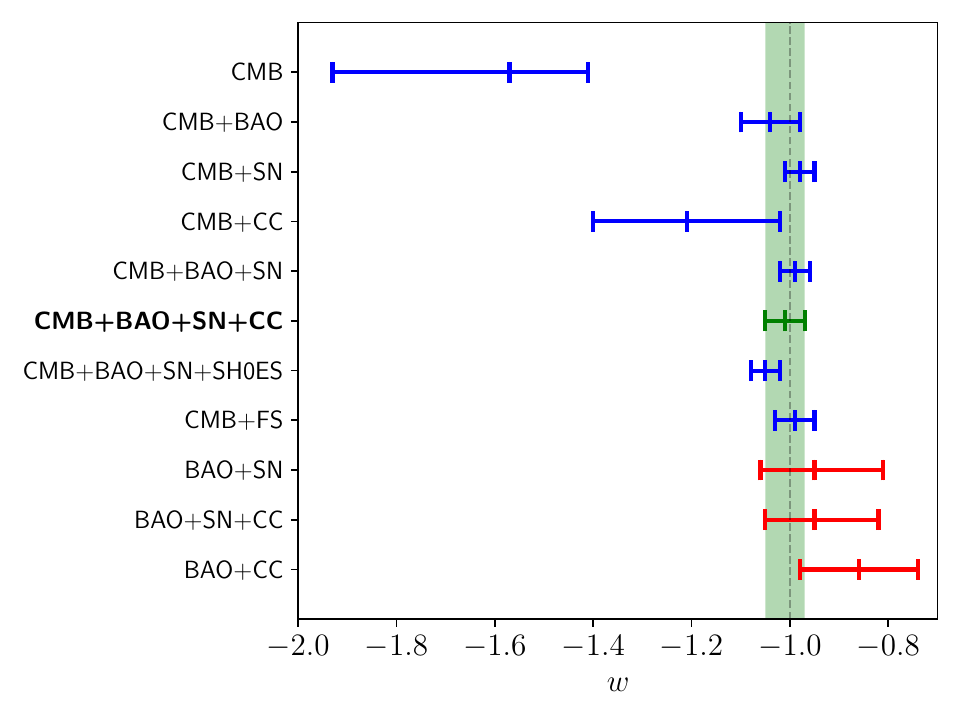}
\caption{Whisker plot summarizing our main results, reporting 68\%~CL intervals on the dark energy equation of state $w$, inferred from a wide variety of dataset combinations assuming the $w$CDM model. Results obtained including \textit{Planck} CMB data are indicated by blue bars (except for the results from our final consensus dataset combination CMB+BAO+SN+CC, indicated by a green bar). whereas results obtained without this dataset are indicated by red bars. The green band denotes the 68\%~CL interval obtained from our final consensus dataset combination CMB+BAO+SN+CC, from which we infer $w=-1.013^{+0.038}_{-0.043}$. Finally, the grey vertical dashed line corresponds to the cosmological constant value, $w=-1$.}
\label{fig:whisker}
\end{figure*}

The overarching goal of this paper has been that of critically examining the state of current constraints on the dark energy equation of state (DE EoS) $w$. We have pursued this task in light of the recent and planned imminent launch of a number of ``Stage IV'' surveys, all of whom share among their main goals that of a precise determination of $w$. Our work has also been partially motivated by the fact that, while broadly consistent with the cosmological constant value $w=-1$, a diverse range of current, independent cosmological probes appear at face value to point towards a slightly phantom DE EoS, $w \sim -1.03$ (see the discussion towards the end of Sec.~\ref{sec:deeos}). With this in mind, our analysis has considered several combinations of state-of-the-art cosmological datasets. Our main results are summarized in \autoref{table:values2}, as well as the whisker plot of \autoref{fig:whisker}, but can be summarized in one sentence: \textit{despite a few scattered hints, we find no compelling evidence forcing us away from the cosmological constant...yet!}. 

We noted (as one can clearly see from \autoref{fig:whisker}) that the central values of $w$ inferred from most dataset combinations involving \textit{Planck} Cosmic Microwave Background (CMB) data tend to fall very slightly within the phantom regime. We have investigated this apparent weak indication for phantom DE in detail, finding no compelling evidence for there being a particular set of datapoints in a particular redshift range driving this result, which we instead have reason to attribute to the CMB dataset itself. When examining the latter, we have not been able to pinpoint any single underlying physical effect driving the apparent preference for phantom DE from CMB data alone (see the uppermost whisker in \autoref{fig:whisker}), which we instead attribute to a combination of effects including the geometrical degeneracy (which in itself is indication that constraints on $w$ from CMB data alone should not be trusted), the late ISW effect and potentially low-$\ell$ anomalies, the lensing anomaly, high-$\ell$ measurements of the damping tail, and constraints on $\tau$ mainly coming from low-$\ell$ E-mode polarization measurements. This interpretation is further supported by the values of $w$ inferred from our CMB-free analyses (see the three red whiskers in \autoref{fig:whisker}). Given the importance of breaking the geometrical degeneracy when deriving constraints on $w$ from combinations involving CMB data, this is the ultimate reason why we have chosen not to accentuate the apparent preference for phantom DE from \textit{Planck} CMB data alone, while emphasizing the importance of combining the latter with late-time datasets. 

Overall, we have judged the combination of CMB, Baryon Acoustic Oscillation, Type Ia Supernovae, and cosmic chronometers data to be particularly trustworthy for what concerns inferences of $w$ (recall in fact that we have used only those cosmic chronometers datapoints for which the full covariance matrix, including the effects of systematics, is available). For this consensus dataset, we find $w=-1.013^{+0.038}_{-0.043}$, in excellent agreement with the cosmological constant value. This is indicated by the green whisker and corresponding green band in \autoref{fig:whisker}.

Where to from here? Despite the overall \textit{broad} agreement with the cosmological constant picture, our opinion is that the community should keep an open mind when deriving constraints on $w$, particularly in the era of tension cosmology. Above all, we believe that the results involving full-shape galaxy clustering measurements, which we touched upon in Sec.~\ref{subsubsec:fs}, deserve further investigation, as they all consistently point towards loosely the same region of phantom DE. We remark once more that if DE were truly phantom, even just \textit{slightly} phantom (say, $w \sim -1.03$), this would force us to rethink a great deal of things from the theory point of view. We plan to investigate these and other related points in future work.

In closing, we note that several of the questions we have left open will hopefully be clarified with data from Stage IV cosmological surveys, some of which have recently launched, whereas others are scheduled for launch soon. Should the uncertainties on $w$ shrink even just by a factor of a few, several of the hints we have reported on throughout the paper could turn into clear evidence for a deviation from the cosmological constant picture, or conversely a resounding confirmation of the latter. We therefore await with excitement these upcoming cosmological observations, which will provide crucial information in the quest towards determining what makes up 70\% of our Universe.

\acknowledgments
\noindent  L.A.E. acknowledges support from the Consejo Nacional de Humanidades, Ciencias y Tecnolog\'{i}as (CONAHCyT, National Council of Humanities Science and Technology of Mexico) and from the Programa de Apoyo a Proyectos de Investigación e Innovación Tecnológica (PAPIIT) from UNAM IN117723. WG acknowledges the support of the E-COST Grant CA21136-27353609 awarded by the European Cooperation in Science and Technology (COST) which funded a short-term scientific mission at the University of Trento primarily dedicated to advancing progress on this research project. E.D.V acknowledges support from the Royal Society through a Royal Society Dorothy Hodgkin Research Fellowship. R.C.N. acknowledges partial support from the Conselho Nacional de Desenvolvimento Cient\'{i}fico e Tecnologico (CNPq, National Council for Scientific and Technological Development) CNPq through project No.~304306/2022-3. S.V.\ acknowledges support from the University of Trento and the Provincia Autonoma di Trento (PAT, Autonomous Province of Trento) through the UniTrento Internal Call for Research 2023 grant ``Searching for Dark Energy off the beaten track'' (DARKTRACK, grant agreement no.\ E63C22000500003), and from the Istituto Nazionale di Fisica Nucleare (INFN) through the Commissione Scientifica Nazionale 4 (CSN4) Iniziativa Specifica ``Quantum Fields in Gravity, Cosmology and Black Holes'' (FLAG). This publication is based upon work from COST Action CA21136 – ``Addressing observational tensions in cosmology with systematics and fundamental physics (CosmoVerse)'', supported by COST (European Cooperation in Science and Technology). We acknowledge the IT Services at the University of Sheffield for the provision of services for High Performance Computing.

\appendix

\section{Impact of the lensing anomaly}
\label{sec:lensing}

\begin{figure}[!ht]
	\centering
	\includegraphics[width=0.8\textwidth]{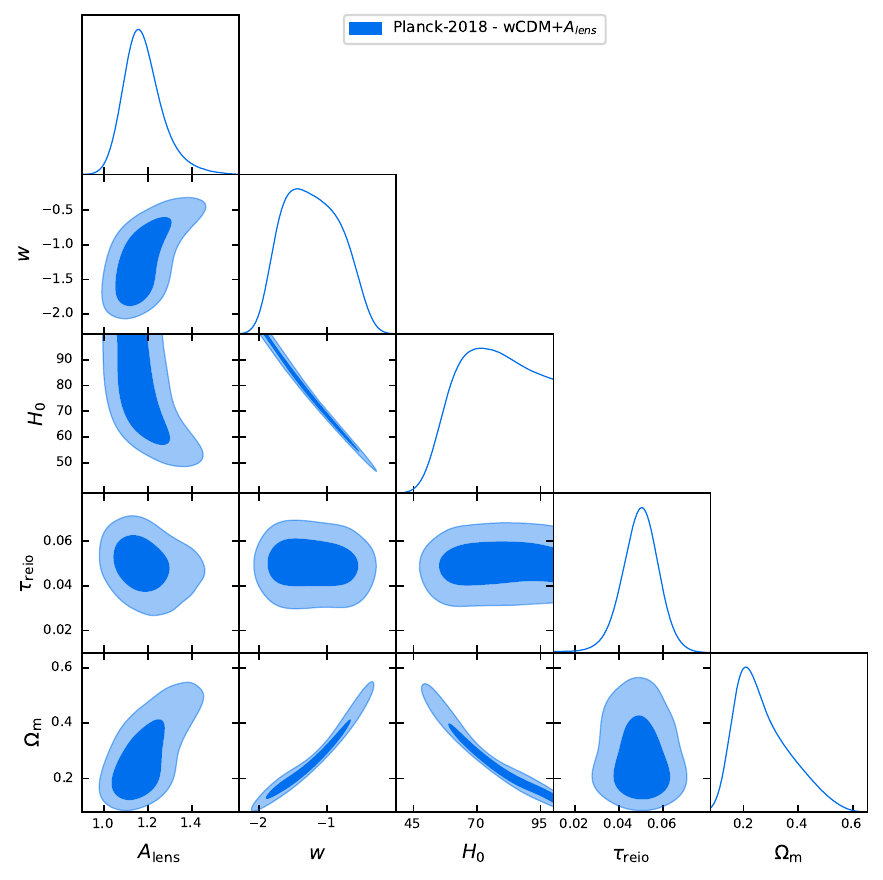}
	\caption{One-dimensional posterior probability distributions and two-dimensional $68\%$ and $95\%$~CL contours for some parameters of interest derived from \textit{Planck} CMB data within the $w$CDM+$A_{\rm lens}$ cosmological model.}
	\label{fig:Alens}
\end{figure}

In recent years, CMB data released by the \textit{Planck} Collaboration have unveiled a few mild anomalies that have become the subject of intense study and debate. One of the most notable issues is the higher lensing amplitude observed in the data, which can be quantified by the phenomenological parameter $A_{\rm lens}$, first introduced in Ref.~\cite{Calabrese:2008rt}. This parameter captures deviations from the lensing amplitude expected within $\Lambda$CDM (corresponding to $A_{\rm lens}=1$). 

When $A_{\rm lens}$ is treated as a free parameter of the cosmological model to be inferred by data, the final analysis of the \textit{Planck} temperature and polarization anisotropies yields the constraint $A_{\rm lens}=1.180 \pm 0.065$, which deviates from the baseline value by about $2.8\sigma$~\cite{Planck:2018vyg}. This anomaly has been well studied and documented in the literature (see e.g.\ Refs.~\cite{DiValentino:2015bja,Renzi:2017cbg,Domenech:2020qay}).

The presence of the lensing anomaly may hold significant implications when constraining other cosmological parameters beyond the standard $\Lambda$CDM model. One notable example pertains to the curvature parameter: since more lensing is expected with a higher abundance of dark matter, the observed lensing anomaly can be recast into a preference for a closed Universe, well documented and discussed in several recent works~\cite{Park:2017xbl,Handley:2019tkm,DiValentino:2019qzk,Efstathiou:2020wem,DiValentino:2020hov,Benisty:2020otr,Vagnozzi:2020rcz,Vagnozzi:2020dfn,DiValentino:2020kpf,Yang:2021hxg,Cao:2021ldv,Dhawan:2021mel,Dinda:2021ffa,Gonzalez:2021ojp,Akarsu:2021max,Cao:2022ugh,Glanville:2022xes,Bel:2022iuf,Yang:2022kho,Stevens:2022evv,Favale:2023lnp}.

In the spirit of testing the potential implications of the lensing anomaly problem for the DE EoS, here we extend our cosmological model, specifically considering the case $w$CDM+$A_{\rm lens}$. We perform a MCMC analysis using the same techniques and methods described in Sec.\ref{sec:data}, while additionally introducing the parameter $A_{\rm lens}$ which is varied within a flat prior range $A_{\rm lens} \in [0, 5]$. The joint constraints on $w$ and $A_{\rm lens}$ obtained solely using \textit{Planck} 2018 CMB data read at 68\% (95\%) CL $w = -1.21 \pm 0.41 (w=-1.21^{+0.85}_{-0.83})$ and $A_{\rm lens} = 1.181^{+0.084}_{-0.10} (A_{\rm lens}=1.18^{+0.20}_{-0.18})$ respectively. 

In \autoref{fig:Alens}, we show the 1D and 2D marginalized contours for various parameters of interest in the analysis, namely $w$, $A_{\rm lens}$, $\tau$, $H_0$ and $\Omega_m$. A few important observations deserve mention: first and foremost, we note that the point ($A_{\rm lens}=1$, $w=-1$), corresponding to $\Lambda$CDM, falls outside the 95\% CL contours. This deviation from the expected values confirms both the mild \textit{Planck} preference for a larger lensing amplitude and a phantom DE EoS. Furthermore, a clear correlation between $A_{\rm lens}$ and $w$ is evident in the \autoref{fig:Alens} and values of the DE EoS close to the cosmological constant $w=-1$ tend to lead to larger values of $A_{\rm lens}>1$. The correlation between these two parameters is, in turn, dependent on the degeneracy among $A_{\rm lens}$, $w$, $H_0$, and $\Omega_m$. Finally, regarding the optical depth at reionization, we observe that when varying $A_{\rm lens}$, $\tau$ shows minimal correlation with both the DE EoS and $A_{\rm lens}$ itself. This finding aligns with the results obtained for $w$CDM when considering the full \textit{Planck} dataset and suggests that the correlations observed in the main part of our work when dividing the \textit{Planck} likelihoods into different bins primarily result from cutting low multipoles and are unlikely to be influenced significantly by the lensing anomaly. Based on these findings, we confidently conclude that the lensing anomaly does not appear to play a predominant role in the interpretation of our results.

\bibliographystyle{apsrev4-1}
\bibliography{wCDM}

\end{document}